\input harvmac
\input epsf

\def \trace{ \mathop{ \rm trace}\nolimits}
\Title{DFTUZ /95-24}
{\vbox{\centerline{Integrable $su(3)$ spin chain combining }
\vskip2pt\centerline{different representations }}}
\centerline{J. Abad and M. R\'{\i}os}
\centerline{Departamento de F\'{\i}sica Te\'{o}rica, Facultad de
Ciencias,} \centerline{Universidad de Zaragoza, 50009 Zaragoza, Spain}
\bigskip
\bigskip
\vskip .3in
\centerline{ \tenbf Abstract}
The general expression for the local matrix $t(\theta)$ of a quantum chain with
the site space in any representation of  $su(3)$ is
obtained.
This is made by generalizing $t(\theta)$ from the fundamental
representation and imposing the fulfillment of the Yang-Baxter
equation.
Then, a non-homogeneous spin chain combining different
representations
of $su(3)$ is solved by developing a method inspired in the nested
Bethe ansatz.
The solution for the eigenvalues of the trace of the monodromy
matrix  is given
as two coupled Bethe equations. A conjecture about the solution of a
chain with
the site states in different representations of $su(n)$ is presented.
The thermodynamic limit of the ground state is calculated.

\bigskip
\bigskip
\Date{}

\vfill
\eject

\newsec{Introduction}

The search for integrable spin chains has deserved considerable
attention in the last years due to the fact that they are interesting physical
systems and have a rich mathematical structure. The best known is the
$XXZ$ Heisenberg $su(2)$ chain with spin $S=1/2$ in every site
\ref\ri{ H. Bethe, Z. Phys. 71 (1931), 205 \semi
R. Orbach, Phys. Rev. 112 (1958), 309\semi
L. R. Walker Phys. Rev. 116 (1959),1089. }, that gave rise to
subsequent development of the quantum groups %
\ref\rnmi{M. Jimbo, Lett. Math. Phys. 10 (1985), 63.}-%
\nref\rnmii{M. Jimbo, Lett. Math. Phys. 11 (1986), 247.}%
\ref\rnmiii{V. G. Drinfeld, Proceedings of the I.C.M. 1986, A. M.
Gleason  editor, (A.M.S. 1987).}.
Integrable spin chains with S=1 and higher spin chains have been
found and solved %
\ref\rv{V. A. Fateev and A.B. Zamolodchikov, Sov. J. Nucl. Phys. 32
(1980),298}%
\nref\rmi{P. P. Kulish and N. Yu Reshetkhin, Soviet Math. 101 (1981),
2435.}%
\nref\rri{H. M. Babujian, Phys. Lett. A 90 (1982) 479.}%
\nref\rrii{L.A. Takhtajan, Phys.Lett. A 87 (1982) 479.}%
\nref\rriii{A. N. Kirillov and N. Y. Reshetikhin, J. Phys. A 20
(1987) 1565.}%
\nref\rriv{H. M. Babujian and A. M. Tsvelick, Nucl. Phys. B 265
(1986) 24.}%
-\ref\rrv{A. B. Zamolodchikov and V. A. Fateev, Sov. J. Nucl. Phys.
32 (1980) 298.}.%
They correspond to higher dimension representations of the quantum
group that give integrable systems of increasing complexity
\ref\rnmiv{V. E. Korepin, N. M. Bogoulibov and A. G. Izergin, Quantum
inverse scatrering method and correlation functions, Cambridge
University Press, Cambridge (1993).}-%
\nref\rnmivd{C. G\'{o}mez, M. Ruiz-Altaba and G. Sierra, Quantum groups in
two-dimensional Physics, Cambridge
University Press, Cambridge (1996). }
\ref\rnmv{J. Abad amd M. R\'{\i}os, J. Phys. A 28 (1995) 3319.}.

In addition, magnetic hamiltonians can be derived from solution of
the Yang-Baxter equations (YBE) %
   (\ref\rnmvi{C. N. Yang, Phys. Rev. Lett. 19 (1967), 1312.}%
\ref\rnmvii{R. J. Baxter, Ann. Phys. 70 (1972), 323.}) %
  associated with Lie algebras other than  $su(2)$
\ref\rmii{H.J. de Vega, J. Mod. Phys. A4 (1989),2371 \semi
H.J. de Vega, Nucl. Phys. B (Proc. and Suppl.) 18A (1990), 229\semi
J. Abad and M. R\'{\i}os, Univ. of Zaragoza Report DFTUZ 94-11
(1994).}. The solutions are found using the Bethe ansatz (BA) for
sites with two components or nested Bethe ansatz (NBA) for sites with
more components %
\ref\rnmviii{B. Sutherland, Phys. Rev. B 12 (1975), 3795.}. The
introduction of the quantum inverse scattering methods (QISM) %
\ref\rnmix{L. D. Faddeev, Sov. Sc. Rev. Math. Phys. C1, (1981), 107.}
gave a systematic method to solve those systems.
The quantum groups give general methods to find new integrable
models.

An interesting problem is to solve integrable chains formed by two
kind of states of the site.  Inhomogeneus solvable models were considered in
\ref\rrvi{P. P. Kulish and N. Yu. Reshetikhin, J. Phys. A 16 (1983), L591.},
(see also \rnmiv ). The simplest case, an alternating chain
with $S=1/2$ and $S=1$ derived from the $su(2)$ Lie algebra was
presented in Ref. \ref\rix{ H.J. de Vega and  F. Woynarovich, J.
Phys. A 25 (1992), 4499.} and in several subsequent works in which
the thermodynamic properties of these systems was studied %
\ref\rx{ H.J. de Vega, L. Mezincescu and R.I. Nepomechie, Phys. Rev.
B 49 (1994), 13223.}%
\nref\rriii{H.J. de Vega, L. Mezincescu and R.I. Nepomechie, J. Mod.
Phys. B8 (1994) 3473.}%
\nref\rriv{M.J. Martins, J. Phys. A 26 (1993) 7301.}%
-\ref\rrv{S. R. Aladin and M.J. Martins, J Phys. A 26 (1993) 1529
and J. Phys A 26
(1993) 7287.}.

The system presents  interesting features; one of them is that it
gives a hamiltonian that contains the usual piece coupling pairs of
neighboring spins $S=1/2$ and $S=1$ and another piece coupling three
neighboring spins. The solution is found using the Bethe ansatz.

In this paper, we are going to solve an alternating chain with the
spin of the sites in the $ \{3\}$ and  $ \{3^*\}$ representations of
$su(3)$.  We have made an extension of  the method used in Ref. \rix\
for systems where the $P$ and $T$ symmetries are not conserved in order to
get hamiltonians associated to alternating chains based on the $su(2)$ algebra.

In a more rigorous sense, we are using the $U_q( su(3))$ algebra and its representations, but can be shown that generally for simple algebras $g$ the representations of $g$ and $U_q( g)$ are isomorphic%
\ref\nui{J. Fuchs, Affine Lie Algebras and Quantum Groups, Cambridge University Press, Cambridge (1992), pag. 267.}.

We can obtain two different systems by using
as auxiliary spaces the representations $\{3\}$ and $\{3^*\}$; they will
give different hamiltonians, but under a relation between the
parameters of the local inhomogeneities that we will specified, we can prove
that they
commute and both systems have the same eigenstates. Then, the
more general system will be a superposition of those two systems.

The diagonalization of these hamiltonians requires important
modifications of the standard method  with the NBA \rmii\
 \rrvi . We start building the monodromy matrix in the auxiliary space whose
elements are operators in the space of states of the chain.
The main  difference is that now we have not a reference state,
eigenstate of the operators in the diagonal of the monodromy matrix and which
is annihilated by all operators under the diagonal of this matrix.
Then, we introduce a reference subspace in the space of states where
we can do the second step of the NBA. So, we obtain the equations of
the ansatz whose properties can be analyzed as in the standard case. The
model, since the auxiliary space has three dimensions, requires only
two steps for the NBA, but the method is easily generalizable to more
dimensions %
\ref \rrvii {J. Abad and M. R\'{\i}os, Phys. Rev. B 53 (1996), 14000.}.

The present paper is organized as follows. In the next section we
develop the technique to obtain the hamiltonians associated to
alternating chains \rix . In the third section we apply the method to
alternating chains with the sites in the $ \{3\}$ and  $ \{3^*\}$
representations of $su(3)$. In the fourth section we find the
eigenvalues of transfer matrix of the system and the equations of our
ansatz as a generalization of the NBA. In section five,  we
analyze of the equations of the ansatz and obtain their thermodynamic limit.

\newsec{Non-homogeneous chain with the site states alternating in two
different representation spaces}
As is well known, regular solutions of the Yang-Baxter equations (YBE)
systematically yield integrable chains. In Ref. \rix\ an integrable quantum
chain
with two types of spins is described.
Following that reference and in order to establish our notation, we
are going to review how an integrable system follows from a $R$-matrix $
R_{c,a}^{b,d} (\theta)$, which is solution of the YBE
\eqn\ebi{ [1\otimes R(\theta-\theta')] [R(\theta)\otimes 1][1\otimes
R(\theta')]=
 [R(\theta')\otimes 1] [1\otimes R(\theta)] [R(\theta-\theta')\otimes
1].}
We associate to each site of the chain the $t$ operator
\eqn\ebii{ [t_{a,b} (\theta)]_{c,d}=R_{c,a}^{b,d} (\theta),}
where the indices $a$ and $b$ act on the site space and the $c$ and
$d$ in an auxiliary space. They are shown graphically in fig. 1 (a).
Then the YBE can be written in the usual form,
\eqn\ebiii{  R(\theta-\theta') \cdot [t(\theta)\otimes  t(\theta')]=
 [t(\theta')\otimes  t(\theta)] \cdot R(\theta-\theta'),}
that graphically is expressed in fig. 2 (a). The $\otimes $ product
is in the site space and the $\cdot$ product is in the auxiliary
space.
\midinsert
\bigskip
\centerline{\epsfxsize=10cm  \epsfbox{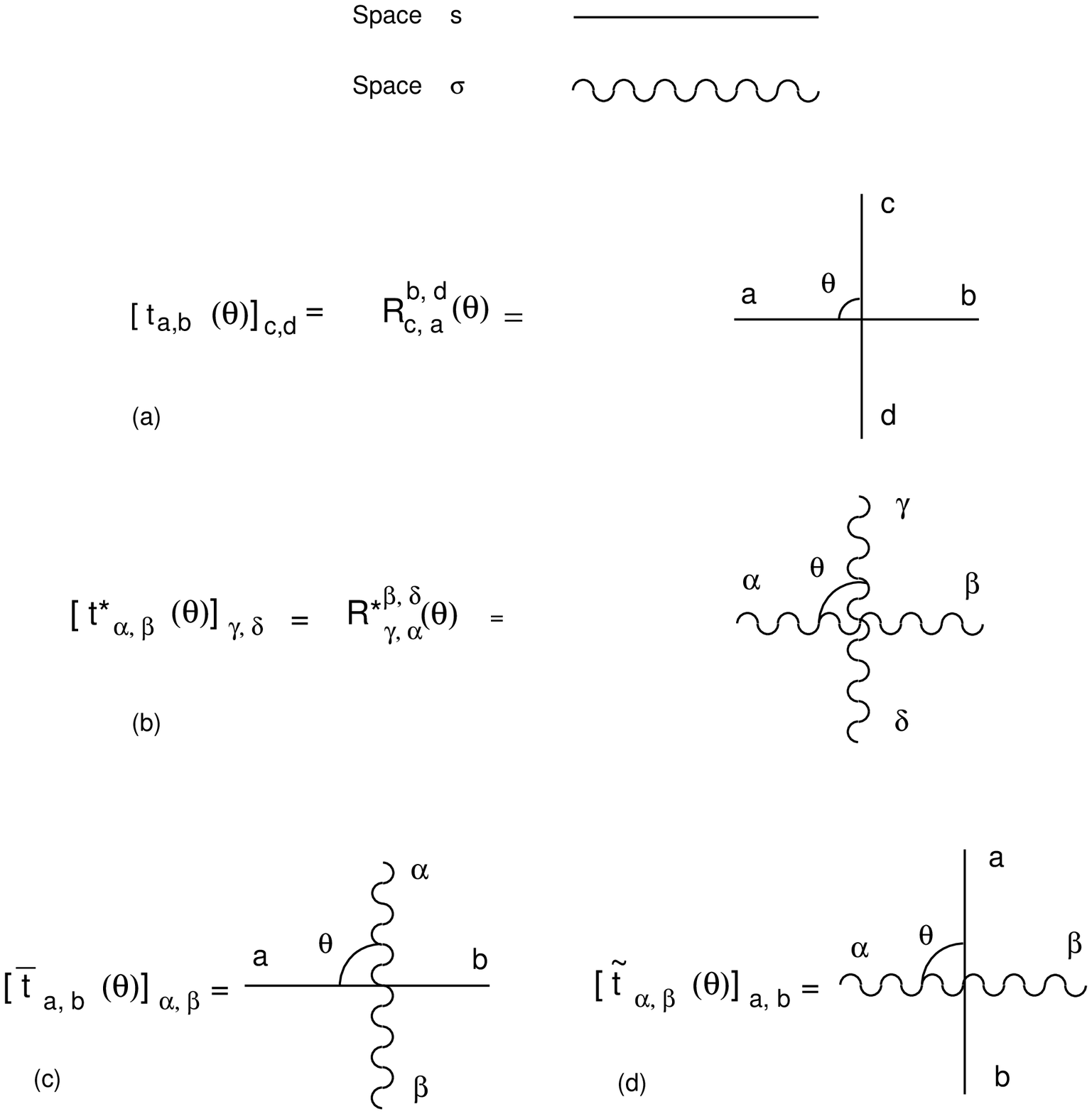}}
\centerline{Figure 1}
\endinsert
\bigskip

Equation \ebi\ is not the most general YBE. In general we have
operators acting on pairs of unequal vector spaces. This is
represented graphically with lines of different kind. We are going to
consider two vector spaces denoted by $s$ and $\sigma$; then we have, besides
 $t$,  the operators $t^* =R^*$, $\bar{t}$ and $\tilde{t}$ represented
in fig. 1. They fulfill the YBEs,
\eqna\ebiv
$$\eqalignno{
& R^*(\theta-\theta') \cdot [t^*(\theta)\otimes  t^*(\theta')]=
 [t^*(\theta')\otimes  t^*(\theta)] \cdot R^*(\theta-\theta'), &\ebiv
a \cr
& R^*(\theta-\theta') \cdot [\tilde{t}(\theta)\otimes
\tilde{t}(\theta')]=
 [\tilde{t}(\theta')\otimes  \tilde{t}(\theta)] \cdot
R^*(\theta-\theta'), &\ebiv b \cr
& R(\theta-\theta') \cdot [\bar{t}(\theta)\otimes  \bar{t}(\theta')]=
 [\bar{t}(\theta')\otimes  \bar{t}(\theta)] \cdot R(\theta-\theta'),
&\ebiv c \cr
}
$$
represented in fig. 2 (b), (c) and (d) respectively.
\midinsert
\bigskip
\centerline{\epsfxsize=10cm  \epsfbox{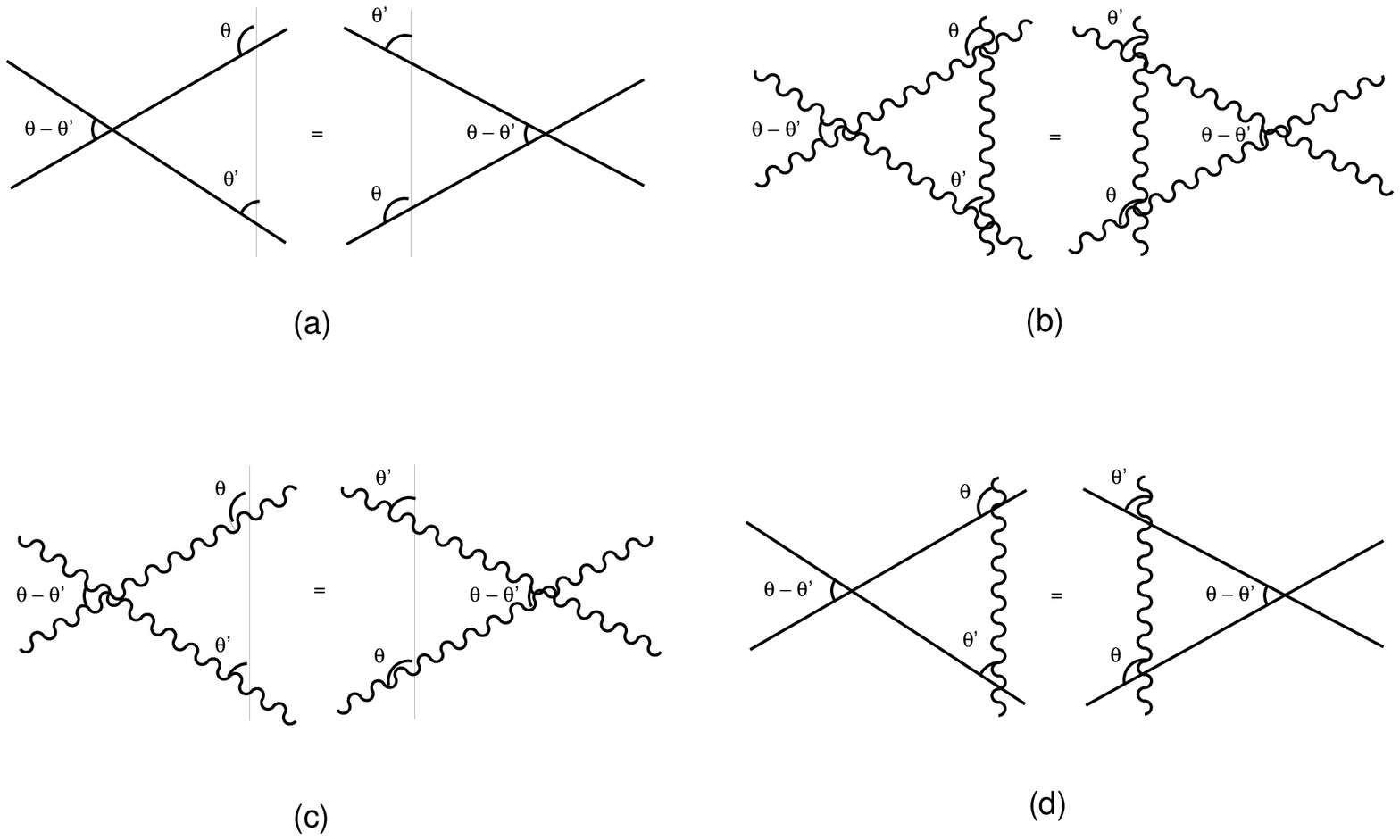}}
\centerline{Figure 2}
\endinsert
\bigskip
In the most general case, we do not require $R(\theta)$ and
$R^*(\theta)$ have $P$ and $T$ symmetry nor to be
invariant under crossing. Instead, we will assume the following properties:

\noindent i) $PT$-symmetry,
\eqna\ebv
$$\eqalignno{R^{c,d}_{a,b}(\theta)&=R^{a,b}_{c,d}(\theta)&\ebv a \cr
R^{*}{}^{\gamma,\delta}_{ \alpha,\beta}(\theta)&=R^{ *}{}^{
\alpha,\beta}_{ \gamma,\delta}
(\theta)&\ebv b \cr}
$$
ii) Unitarity,
\eqna\ebvi
$$\eqalignno{&R^{c,d}_{a,b}(\theta) R^{e,f}_{c,d}(-\theta) =
\rho(\theta) \delta_{a,e} \delta_{b,f} ,&\ebvi a \cr
&R^{*}{}^{\gamma,\delta}_{ \alpha,\beta}(\theta) R^{*}{}^{\mu,\nu}_{
\gamma,\delta}(-\theta)=\rho^*(\theta) \delta_{\alpha,\mu}
\delta_{\beta,\nu} ,&\ebvi b \cr}
$$
iii) Regularity,
\eqn\ebvii
{
R(0)= c_0 I ,
}
iv)A matrix $M$ exits such that
\eqn\ebviii{
R^{c,d}_{a,b}(\theta) M_{b,e} R^{g,e}_{f,d}(-\theta-2\eta)
M^{-1}_{f,h}\propto
\delta_{a,g}\delta_{c,h},
}
v) The $t$-matrices verify,
\eqn\ebix{
[\bar{t}_{a,b}(\theta)]_{\alpha,\beta}  [
\tilde{t}_{\beta,\gamma}(-\theta)]_{b,c}=\tilde{\rho} (\theta)
\delta_{a,c}\delta_{\alpha,\gamma}.
}

We consider a non-homogeneous chain with  $2 N$ sites in which the
site spaces are alternating in the representations $\{3\}$ and
$\{3^*\}$. This chain has associated the operator
\eqn\ebx{
T^{\rm{(alt)}}_{a,b} (\theta,\alpha)= t^{(1)}_{a,a_1}(\theta)\bar{t}
^{(2)}_{a_1,a_2}
(\theta+\alpha) \ldots
t^{(2N-1)}_{a_{2N-2},a_{2N-1}}(\theta)\bar{t} ^{(2N)}_{a_{2N-1,b}}
(\theta+\alpha)
}
which is a matrix in the auxiliary space called monodromy matrix,
since it describes the transportation along the chain. The elements
of this matrix are operators on the space tensor product of the site
spaces. It is graphically represented by fig. 3.
\midinsert
\bigskip
\centerline{\epsfxsize=10cm  \epsfbox{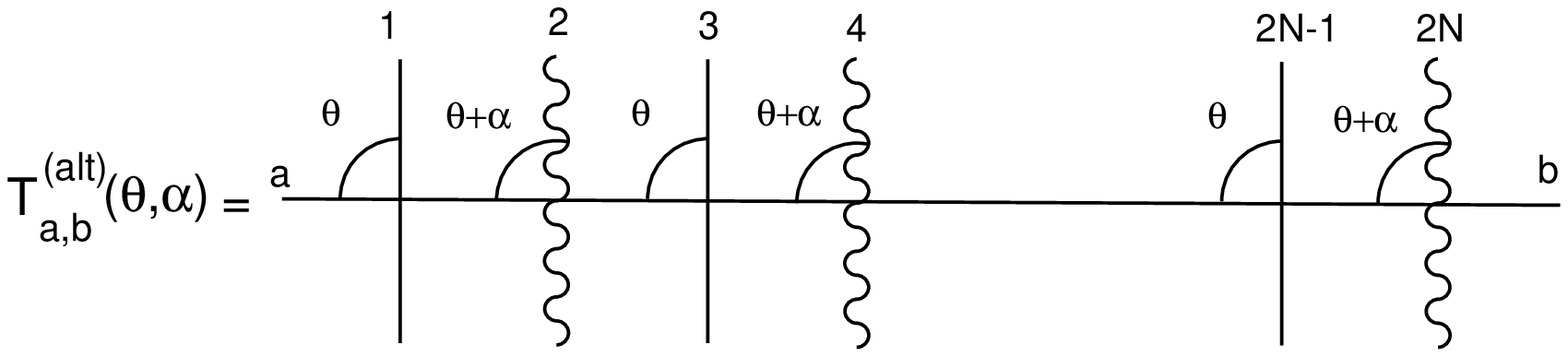}}
\centerline{Figure 3}
\endinsert
\bigskip
%
%

Since the $t$ and $\bar{t}$ matrices fulfill \ebiii , the $T^{\rm{(alt)}}$ also
verifies the YBE
\eqn\ebxi{  R(\theta-\theta') [T^{\rm{(alt)}}(\theta)\otimes
T^{\rm{(alt)}}(\theta')]=
 [T^{\rm{(alt)}}(\theta')\otimes  T^{\rm{(alt)}}(\theta)]
R(\theta-\theta').}
Following the standard procedure, we take the transfer matrices
\eqn\ebxii{\tau^{\rm{(alt)}}(\theta,\alpha) =
T^{\rm{(alt)}}_{a,a}(\theta,\alpha).}
which are the trace of the monodromy matrices. Due to \ebxi {\bf }, the
operators corresponding to different values of the argument  $\theta$ do
commute,
\eqn\ebxiii{[\tau^{\rm{(alt)}}(\theta,\alpha),
\tau^{\rm{(alt)}}(\theta',\alpha)] = 0  .}

The successive derivatives of the transfer matrix at $\theta=0$ give us a
family of commuting operators that describe a solvable system,
the hamiltonian of that system being the first derivative,
\eqn\ebxiv{
H={d \over
d\theta}{\ln{\tau^{\rm{(alt)}}(\theta,\alpha)}\bigl|_{\theta=0}} .
}

In a homogeneous chain the hamiltonian is a sum of nearest
neighbor interactions terms (two-site operators). In our case, it is
very different due to inhomogeneities and there are also
next-to-nearest neighbor interaction terms (three-site operators).
 Collecting separately the two kinds of terms, the hamiltonian becomes
\eqn\ebxv{
H={1 \over{\tilde{\rho}(\alpha)}}
\sum_{i=1 \atop i={\rm odd}}^{2N-1}{h^{(1)}_{i, i+1}} +
{1 \over{c_0 \tilde{\rho}(\alpha)}}
\sum_{i=1 \atop i ={\rm odd}}^{2N-1}{h^{(2)}_{i, i+1,i+2}} ,
}
with
\eqn\ebxvi{
(h^{(1)}_{i, i+1})_{a, \beta;   b, \gamma}=
[\dot{\bar{t}} _{a, c}(\alpha)]_{\beta,  \delta}
                      [ \tilde{t}_{\delta, \gamma}(-\alpha)]_{c,b} ,
}
and
\eqn\ebxvii{
(h^{(2)}_{i, i+1, i+2})_{a, \beta, c;   b, \gamma, d}=
[\bar{t} _{a, e}(\alpha)]_{\beta,  \delta}
[ \dot{t}_{e, d}(0) ]_{c, f}
[ \tilde{t}_{\delta, \gamma}(-\alpha)]_{f,b}  ,
}
that graphically are expressed in fig. (4.a) and (4.b) respectively.
\midinsert
\bigskip
\centerline{\epsfxsize=10cm  \epsfbox{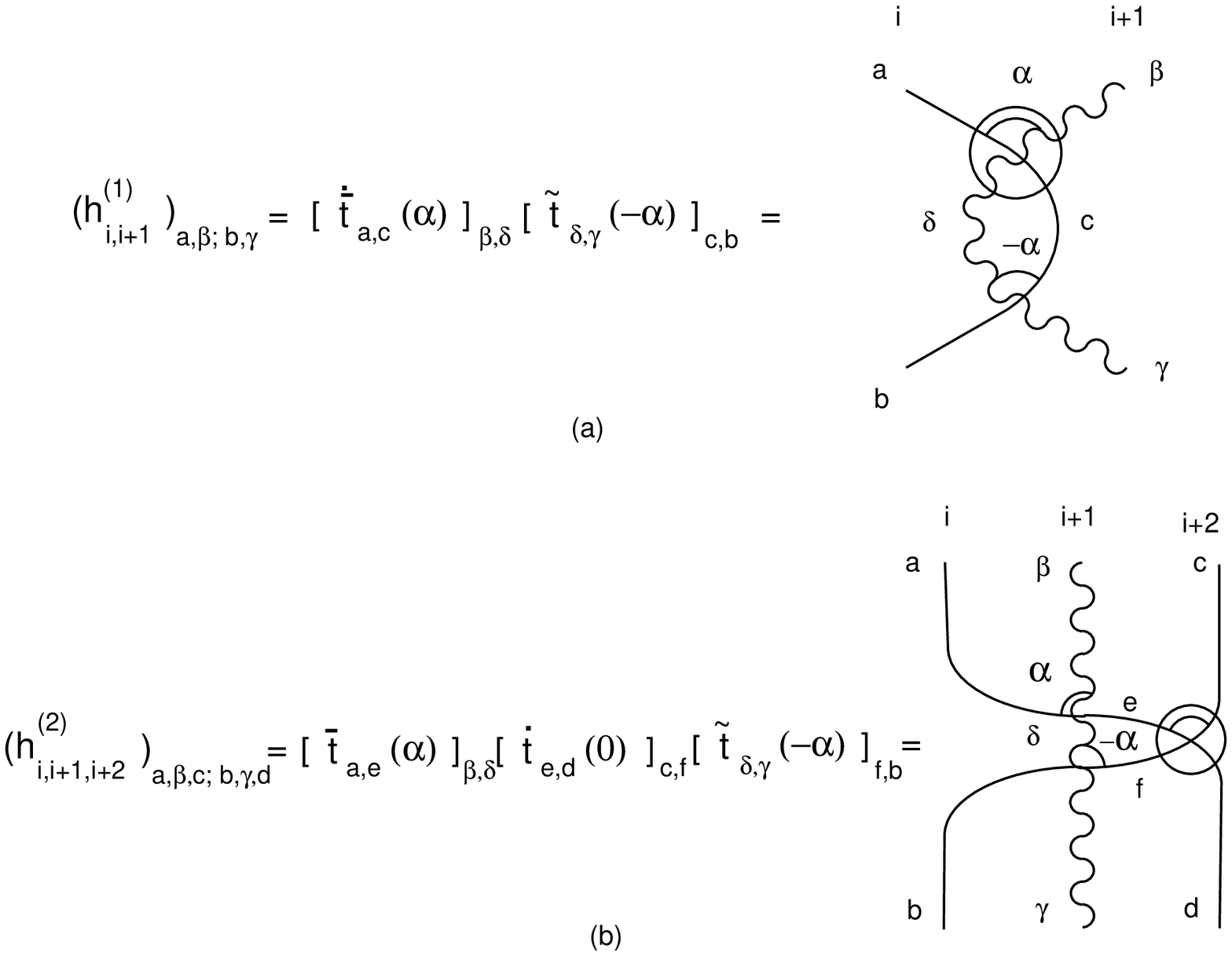}}
\centerline{Figure 4}
\endinsert
\bigskip
A similar process can be made by using as auxiliary space the
$\sigma$ one. Thus, we define the new monodromy matrix
\eqn\ebxviii{
\tilde{T}^{\rm{(alt)}}_{\alpha, \beta} (\theta,\sigma)=
\tilde{t}^{(1)}_{\alpha,\alpha_1}(\theta+\sigma){t} ^{* (2)
}_{\alpha_1,\alpha_2}
(\theta) \ldots
\tilde{t}^{(2N-1)}_{\alpha_{2N-2},\alpha_{2N-1}}(\theta+\sigma){t}
^{*(2N) }_{\alpha_{2N-1,\beta}}
(\theta) ,
}
graphically represented in fig. 5. It fulfills the YBE
\eqn\ebxix{
R^*(\theta-\theta')[\tilde{T}^{\rm{(alt)}}(\theta
-\sigma)\otimes\tilde{T}^{\rm{(alt)}}(\theta'-\sigma)]=
[\tilde{T}^{\rm{(alt)}}(\theta'-
\sigma)\otimes\tilde{T}^{\rm{(alt)}}(\theta-\sigma)]R^*(\theta-\theta') .
}
\midinsert
\bigskip
\centerline{\epsfxsize=10cm  \epsfbox{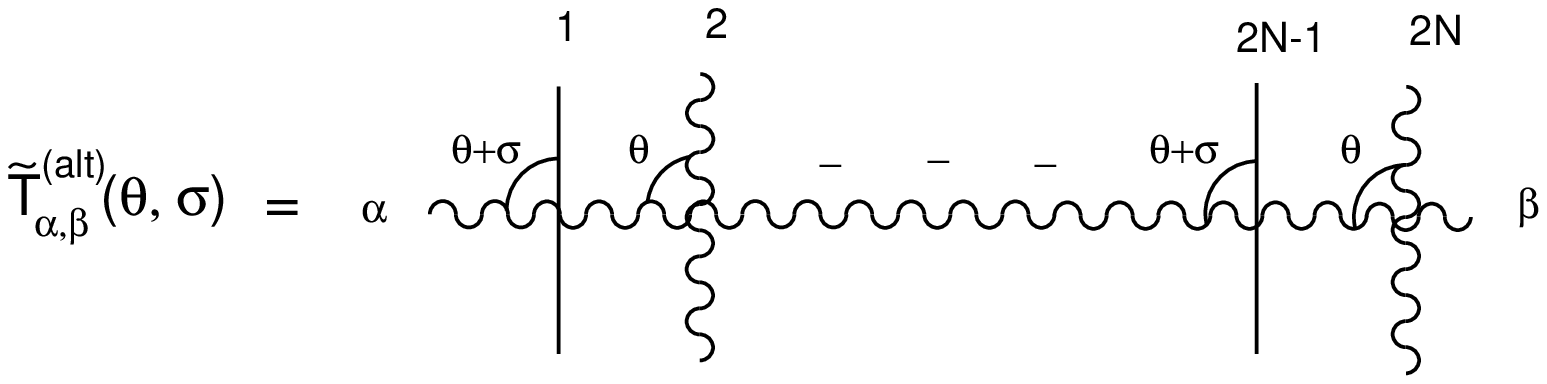}}
\centerline{Figure 5}
\endinsert
\bigskip
The hamiltonian obtained from this monodromy matrix using a
formula similar to \ebxii\ is
\eqn\ebxx{
\tilde{H}={1 \over{\tilde{\rho}(\sigma)}}
\sum_{i=2 \atop i={\rm  even}}^{2N}{\tilde{h}^{(1)}_{i, i+1}} +
{1 \over{c_0 \tilde{\rho}(\sigma)}}
\sum_{i=2 \atop i ={\rm even}}^{2N}{\tilde{h}^{(2)}_{i, i+1,i+2}} ,
}
with
\eqn\exxi{
(\tilde{h}^{(1)}_{i,i+1})_{\alpha, a; \beta, b}=
[\dot{\tilde{t}}_{\alpha,\delta}(\sigma)]_{a,c}
[\bar{t}_{c,b}(-\sigma)]_{\delta,\beta} ,
}
and
\eqn\exxii{
(\tilde{h}^{(2)}_{i,i+1, i+2})_{\alpha, a,\mu; \beta, b,\nu}=
[\tilde{t}_{\alpha,\delta}(\sigma)]_{a,c}
[\dot{t}^*_{\delta,\nu}(0)]_{\mu,\rho}
[\bar{t}_{c,b}(-\sigma)]_{\rho,\beta} .
}

The monodromy matrices $ T^{\rm{(alt)}} $ and $
\tilde{T}^{\rm{(alt)}} $ fulfill the following YBE
\eqn\ebxxiii{
[\bar{t}_{a,b}(\theta-\theta' +\gamma)]_{\alpha,\beta}
T^{\rm{(alt)}}_{b,c} (\theta,\gamma)
\tilde{T}^{\rm{(alt)}}_{\beta,\delta}(\theta', -\gamma)=
\tilde{T}^{\rm{(alt)}}_{\alpha,\mu}(\theta', -\gamma)
T^{\rm{(alt)}}_{a, d} (\theta,\gamma)
[\bar{t}_{d, c}(\theta-\theta' +\gamma)]_{\mu, \delta} ,
}
that graphically is expressed in fig. 6.
\midinsert
\bigskip
\centerline{\epsfxsize=10cm  \epsfbox{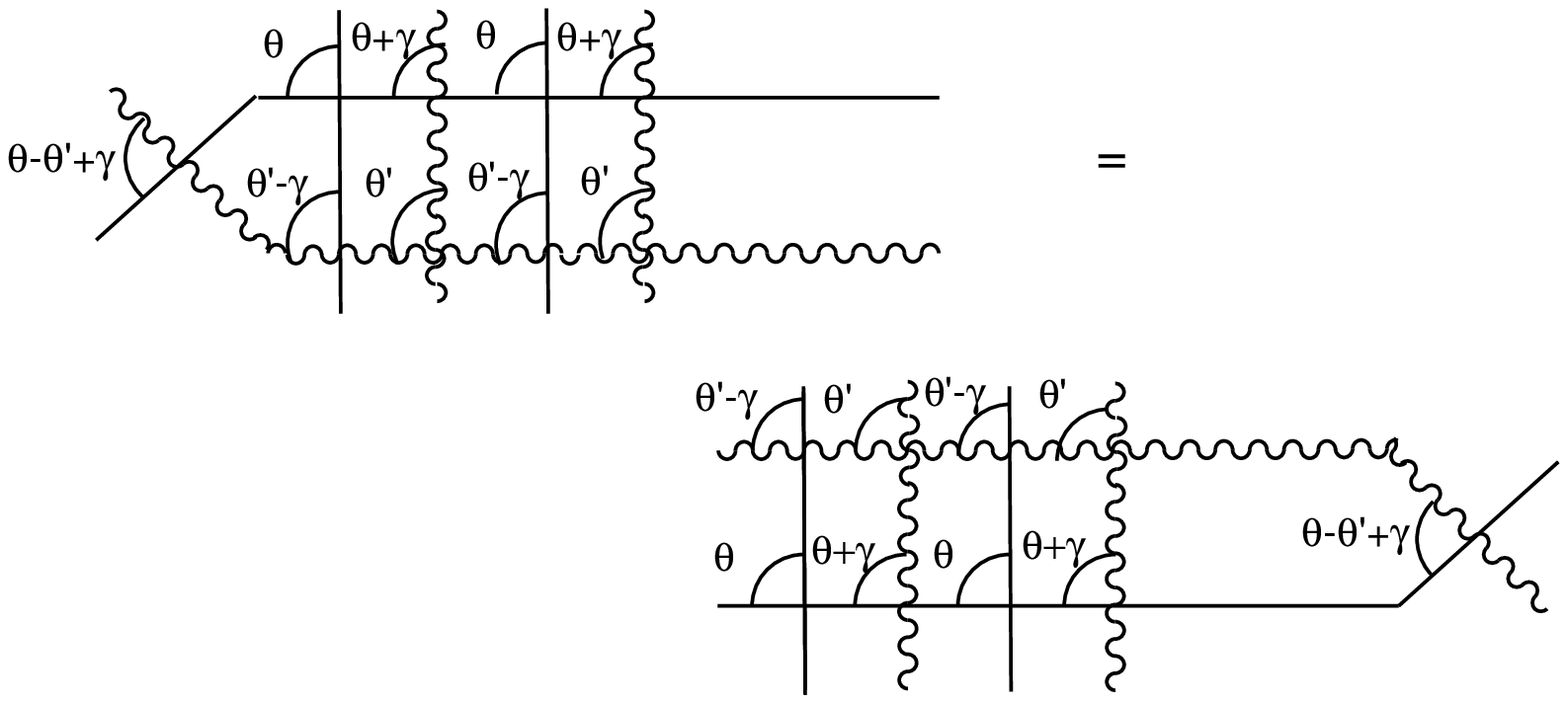}}
\centerline{Figure 6}
\endinsert
\bigskip
As a consequence of \ebxxiii\ the transfer matrices
$\tau^{\rm{(alt)}}$ and $\tilde{\tau}^{\rm{(alt)}}$ commute
\eqn\ebxxiv{
[ \tau^{\rm{(alt)}}(\theta, \alpha),
\tilde{\tau}^{\rm{(alt)}}(\theta', - \alpha)]=0  ,
}
and then, the derived hamiltonians $H$ and $\tilde{H}$ also commute.
Thus, both can be simultaneously diagonalized with common eigenstates.

\newsec{Quantum chain with the site states alternating in two
representations of $su(3)$}
We describe in this section a non-homogeneous chain that we form
alternating two representations of $su(3)$. We denote a representation
by the indices of its associated Dynkin diagram $(m_1, m_2)$.
The vector space $s$ is taken as the representation $(1, 0) \equiv
\{3\}$ and the space $\sigma$ is the generic representation $(m_1,
m_2)$.

The $t$ operator acting on the site and auxiliary spaces, both $s$, \rrvii
can be written \rrvii\
\eqn\eci{
t(\theta,\gamma)=
\pmatrix{
{1 \over 2}(\lambda^{3}q^{-N^{\alpha}}-\lambda^{-3}q^{N^{\alpha}}) &
\lambda
{(q^{-1}-q)
\over 2} f_{1} & \lambda^{-1}{(q^{-1}-q) \over 2} [f_{2},f_{1}] \cr
\lambda^{-1}{(q^{-1}-q) \over 2} e_{1} &{1 \over
2}(\lambda^{3}q^{-N^{\beta}}-
\lambda^{-3}q^{N^{\beta}}) & \lambda {(q^{-1}-q)\over 2} f_{2} \cr
\lambda{(q^{-1}-q) \over 2} [e_{1},e_{2}]  & \lambda^{-1}{(q^{-1}-q)
\over 2}
e_{2} &
{1 \over 2}(\lambda^{3}q^{-N^{\gamma}}-\lambda^{-3}q^{N^{\gamma}}) \cr
},}
where the parameters  $\lambda$ and $q$ have been taken as the
functions of $\theta$ and $\gamma$
\eqn\ecii{\lambda = e^{\theta \over 2}, \qquad q = e^{-\gamma},}
and the $N$ matrices are
\eqna\eciii
$$\eqalignno{N^{\alpha}&={2 \over 3}h_{1}+{1 \over 3}h_{2}+{1 \over
3}I,&\eciii a
\cr
N^{\beta}&=-{1 \over 3}h_{1}+{1 \over 3}h_{2}+{1 \over 3}I,&\eciii b
\cr
N^{\gamma}&=-{1 \over 3}h_{1}-{2 \over 3}h_{2}+{1 \over 3}I,&\eciii c
\cr}
$$
where $\{e_{i}, f_{i}, q^{\pm h_{i}}\}$, $i=1, 2$, are the Cartan
generators of the
deformed algebra $U_{q}(sl(3))$.

To obtain the operators ${\bar{t}}(\theta,\gamma) $,
we take \eci\ as a basis and write
\eqn\eciv{
\bar{t}(\lambda)=
\pmatrix{
{1 \over 2}(\lambda^{3}q^{-N^{\alpha}}-\lambda^{-3}q^{N^{\alpha}}) &
\lambda
{(q^{-1}-q)
\over 2} F_{1} & \lambda^{-1}{(q^{-1}-q) \over 2} F_{3} \cr
\lambda^{-1}{(q^{-1}-q) \over 2} E_{1} &{1 \over
2}(\lambda^{3}q^{-N^{\beta}}-
\lambda^{-3}q^{N^{\beta}}) & \lambda {(q^{-1}-q)\over 2} F_{2} \cr
\lambda{(q^{-1}-q) \over 2} E_{3}  & \lambda^{-1}{(q^{-1}-q) \over 2}
E_{2} &
{1 \over 2}(\lambda^{3}q^{-N^{\gamma}}-\lambda^{-3}q^{N^{\gamma}}) \cr
},}
where the operators $\{E_{i},F_{i}\}$, $i=1,3$, are  unknown and will
be
determined by imposing the YBE,
\eqn\ecv {R(\theta-\theta', \gamma) \cdot [\bar{t}(\theta,
\gamma)\otimes  \bar{t}(\theta',  \gamma)]=
 [\bar{t}(\theta',  \gamma)\otimes  \bar{t}(\theta,  \gamma)] \cdot
R(\theta-\theta',  \gamma), }
that is shown in fig. 2(d).
The $R_{c, a} ^ {b,d}(\theta)\equiv [ t_{a,b} (\theta,\gamma)]_{c,d}$
is given by
\eqn\ecvi{
R(\lambda,\mu)=\left (\matrix{
a & 0 & 0 & 0 & 0 & 0 & 0 & 0 & 0 \cr
0 & d & 0 & b & 0 & 0 & 0 & 0 & 0 \cr
0 & 0 & c & 0 & 0 & 0 & b & 0 & 0 \cr
0 & b & 0 & c & 0 & 0 & 0 & 0 & 0 \cr
0 & 0 & 0 & 0 & a & 0 & 0 & 0 & 0 \cr
0 & 0 & 0 & 0 & 0 & d & 0 & b & 0 \cr
0 & 0 & b & 0 & 0 & 0 & d & 0 & 0 \cr
0 & 0 & 0 & 0 & 0 & b & 0 & c & 0 \cr
0 & 0 & 0 & 0 & 0 & 0 & 0 & 0 & a \cr
}\right ),
}
with
\eqna\ecvii
$$\eqalignno{a(\lambda,\mu)&={1 \over 2}(\lambda^{3}\mu^{-3}q^{-1} -
\lambda^{-3}\mu^{3}q) ,&\ecvii a  \cr
b(\lambda,\mu)&={1 \over 2}(\lambda^{3}\mu^{-3} -
\lambda^{-3}\mu^{3}) ,&\ecvii b \cr
c(\lambda,\mu)&={1 \over 2}(q^{-1}-q)\lambda\mu^{-1} ,&\ecvii c \cr
d(\lambda,\mu)&={1 \over 2}(q^{-1}-q)\lambda^{-1}\mu .&\ecvii d \cr}
$$

The relations obtained are
\eqna\ecviii
$$\eqalignno{E_{1}q^{N^{\alpha}} &= q^{-1}q^{N^{\alpha}} E_{1},
&\ecviii a  \cr
E_{1}q^{N^{\beta}} &= qq^{N^{\beta}} E_{1}, &\ecviii b \cr
F_{1}q^{N^{\alpha}} &= qq^{N^{\alpha}} F_{1}, &\ecviii c \cr
F_{1}q^{N^{\beta}} &= q^{-1} q^{N^{\beta}} F_{1}, &\ecviii d \cr
E_{2}q^{N^{\alpha}} &= qq^{N^{\alpha}} E_{2}, &\ecviii e \cr
E_{2}q^{N^{\beta}} &= q^{-1}q^{N^{\beta}} E_{2}, &\ecviii f \cr
F_{2}q^{N^{\alpha}} &= q^{-1}q^{N^{\alpha}} F_{2},&\ecviii g \cr
F_{2}q^{N^{\beta}} &= qq^{N^{\beta}} F_{2} ,&\ecviii h \cr
[E_{1},F_{1}]&=(q^{-1}-q)\left(q^{N^{\beta}-N^{\alpha}} -
q^{N^{\alpha}-N^{\beta}} \right), &\ecviii i \cr
[E_{2},F_{2}]&=(q^{-1}-q)\left(q^{N^{\gamma}-N^{\beta}} -
q^{N^{\beta}-N^{\gamma}} \right), &\ecviii j \cr
E_{3} &= {1 \over (q^{-1}-q)}q^{-N^{\beta}}[E_{1},E_{2}] ,&\ecviii k
\cr
F_{3} &= {1 \over (q^{-1}-q)}q^{N^{\beta}}[F_{2},F_{1}] ,&\ecviii l
\cr}
$$
and besides, the modified Serre relations
\eqna\ecix
$$\eqalignno{&q^{-1}E_{1}E_{1}E_{2}-(q+q^{-1})E_{1}E_{2}E_{1} +
qE_{2}E_{1}E_{1} =0, &\ecix a  \cr
&qE_{2}E_{2}E_{1}-(q+q^{-1})E_{2}E_{1}E_{2} +
q^{-1}E_{1}E_{2}E_{2} = 0, &\ecix b \cr
&q^{-1}F_{1}F_{1}F_{2}-(q+q^{-1})F_{1}F_{2}F_{1} +
qF_{2}F_{1}F_{1} = 0, &\ecix c \cr
&qF_{2}F_{2}F_{1}-(q+q^{-1})F_{2}F_{1}F_{2} +
q^{-1}F_{1}F_{2}F_{2} = 0 ,&\ecix d \cr}
$$
should be verified. It must be noted that that the relations
 \ecviii {a\hbox{--}l}\  are
the usual ones for the quantum group
$U_{q}(sl(3))$ while the relations \ecix {a\hbox{--} d}\ are not the
usual ones for the said group, and
because of this, the YBE is not verified if the
generators  $e_i$ and $f_i$, pertaining to the deformed algebra, are
taken as $E_i$ and $F_i$. This induces us to take
\eqna\ecx
$$\eqalignno{F_{i}&={1 \over 2}(q^{-1}-q)Z_{i}f_{i}, &\ecx a  \cr
E_{i}&={1 \over 2}(q^{-1}-q)e_{i}Z_{i}^{-1} ,\qquad \qquad i=1,2
&\ecx b \cr}
$$
where $e_i$ and $f_i$, $i=1,2$, are the generators of $U_{q}(sl(3))$
in the
representation $(m_1 , m_2)$ and $Z_i$ are two diagonal operators
that were obtain by imposing the verification of the relations
\ecviii  {a\hbox{--}l}\ and \ecix {a\hbox{--}d}\ . In
this way, one obtains the general form of these operators given by
\eqna\ecxi
$$\eqalignno{Z_{1}&=q^{a_{1}h_{1}-{1 \over 3}h_{2}+a_{3}I}, &\ecxi a
\cr
Z_{2}&=q^{{1 \over 3}h_{1}+(a_{1}+{1 \over 3})h_{2}+b_{3}I}. &\ecxi b
\cr}
$$
The knowledge of the operator $\bar{t}$ permits us to build the
monodromy operator of any
multistate  chain that mixes two representations. As an example, for
the chain that mixes the $\{3\}$ and the $(m_1, m_2)$ representations
the monodromy operator is%
\eqn\ecxii{
T^{\rm (alt)}_{a,b} (\theta)= t^{(1)}_{a,a_1}(\theta) {\bar{t}}
^{(2)}_{a_1,a_2}
(\theta) \ldots
t^{(2N-1)}_{a_{2N-2},a_{2N-1}}(\theta){\bar{t} }^{(2N)}_{a_{2N-1,b}}
(\theta) ,
}
that is represented graphically as shown in fig. 3.

\newsec{Bethe ansatz equations of the models with space sites in
different representations of $su(3)$}
In this section, we are going to solve an alternating chain that
mixes the $\{3\}$ and $\{3^*\}$ representations of $su(3)$ and the
results will be generalized to chains that mix two arbitrary
representations.

In this case, the $t $ operator is given by \eci\ that can be written
in matrix form
\eqn\edi{
t(\theta)=
\left (\matrix{
a & 0 & 0 & 0 & 0 & 0 & 0 & 0 & 0 \cr
0 & b & 0 & c & 0 & 0 & 0 & 0 & 0 \cr
0 & 0 & b & 0 & 0 & 0 & d & 0 & 0 \cr
0 & d & 0 & b & 0 & 0 & 0 & 0 & 0 \cr
0 & 0 & 0 & 0 & a & 0 & 0 & 0 & 0 \cr
0 & 0 & 0 & 0 & 0 & b & 0 & c & 0 \cr
0 & 0 & c & 0 & 0 & 0 & b & 0 & 0 \cr
0 & 0 & 0 & 0 & 0 & d & 0 & b & 0 \cr
0 & 0 & 0 & 0 & 0 & 0 & 0 & 0 & a \cr
}\right ),
}
with
\eqna\edii{
$$\eqalignno{
&a(\theta)=\sinh{({3 \over 2} \theta+\gamma)},&\edii a\cr
&b(\theta)= \sinh{({3 \over 2} \theta)} ,&\edii b \cr
&c(\theta)=\sinh{(\gamma)}  e^{{ \theta\over 2}},&\edii c \cr
&d(\theta)=\sinh{(\gamma)} e^{{ -\theta\over 2}}.&\edii d
\cr
}
$$}

In the same way, the $\bar{t}$ is obtained from \eciv\ by taking in
\ecx {a, b}\ the  generators of $su(3)$ in the $\{3^*\}$ representation,
\eqn\ediii{
e_1=\pmatrix{
0 & 0 & 0 \cr
-1 & 0 & 0 \cr
0 & 0 & 0 \cr
},
e_2=\pmatrix{
0 & 0 & 0 \cr
0 & 0 & 0 \cr
0 & -1& 0 \cr
},
f_1=\pmatrix{
0 & -1& 0 \cr
0 & 0 & 0 \cr
0 & 0 & 0 \cr
},
f_2=
\pmatrix{
0 & 0 & 0 \cr
0 & 0 & -1 \cr
0 & 0 & 0 \cr
}.
}

Besides, we must fix in \ecxi{a, b}\ the values of $ a_1$, $a_3$ and $b_3$.
By taking
\eqn\ediv{
 a_1= {2 \over 9}, \qquad  a_3= 0, \qquad b_3=0,
}
and rescaling $\theta$ by
\eqn\edv{
\theta= \theta + {5 \over 9} \gamma,
}
we find
\eqn\edvi{
\bar{t}(\theta)=
\left (\matrix{
\bar{a }& 0 & 0 & 0& \bar{c}& 0 & 0 & 0 &  \bar{d } \cr
0 &\bar{ b} & 0 & 0 & 0 & 0 & 0 & 0 & 0 \cr
0 & 0 & \bar{ b} & 0 & 0 & 0 & 0 & 0 & 0 \cr
0 & 0 & 0 &\bar{ b} & 0 & 0 & 0 & 0 & 0 \cr
\bar{d } & 0 & 0 & 0 & \bar{a} & 0 & 0 & 0 & \bar{c} \cr
0 & 0 & 0 & 0 & 0 & \bar{b }& 0 & 0 & 0 \cr
0 & 0 & 0 & 0 & 0 & 0 & \bar{b} & 0 & 0 \cr
0 & 0 & 0 & 0 & 0 & 0 & 0 &\bar{ b} & 0 \cr
\bar{c} & 0 & 0 & 0 &  \bar{d } & 0 & 0 & 0 &\bar{ a }\cr
}\right ),
}
with
\eqna\edvii{
$$\eqalignno{
&\bar{a}(\theta)=\sinh{({3 \over 2} \theta+{\gamma \over 2})},&\edvii
a\cr
&\bar{b}(\theta)= \sinh{({3 \over 2} (\theta+\gamma))} ,&\edvii b \cr
&\bar{c}(\theta)=-\sinh{(\gamma)}  e^{{ (\theta+\gamma)\over
2}},&\edvii c \cr
&\bar{d}(\theta)=-\sinh{(\gamma)} e^{{ -(\theta+\gamma)\over
2}}.&\edvii d
\cr
}
$$}
As we take the $\{3^*\}$ representation as auxiliary space, the $R$
matrix is
\eqn\edviii{
R(\theta)=
\left (\matrix{
a & 0 & 0 & 0 & 0 & 0 & 0 & 0 & 0 \cr
0 & d & 0 & b & 0 & 0 & 0 & 0 & 0 \cr
0 & 0 & c & 0 & 0 & 0 & b & 0 & 0 \cr
0 & b & 0 & c & 0 & 0 & 0 & 0 & 0 \cr
0 & 0 & 0 & 0 & a & 0 & 0 & 0 & 0 \cr
0 & 0 & 0 & 0 & 0 & d & 0 & b & 0 \cr
0 & 0 & b & 0 & 0 & 0 & d & 0 & 0 \cr
0 & 0 & 0 & 0 & 0 & b & 0 & c & 0 \cr
0 & 0 & 0 & 0 & 0 & 0 & 0 & 0 & a \cr
}\right )
}
and in this case the matrix $M$ defined in \ebviii\ is the identity
and the functions defined in the equations \ebvi{a, b}\ and \ebvii\ are
\eqna\edix{
$$\eqalignno{
&\rho(\theta) = \rho^*(\theta)=\sinh{(\gamma-{3 \over 2}\theta)}
\sinh{(\gamma+{3 \over 2}\theta)} , &\edix a \cr
&c_0=c_0^*=\sinh{\gamma}, &\edix b \cr
&\tilde{\rho}(\theta) ={1 \over 2} \left( \cosh{(3\gamma)} -
\cosh{(3\theta)}\right) . &\edix c \cr
}
$$}
We group two neighbor sites in the chain and form the operator
\eqn\edx{
\hat{t}^{(i, i+1)}_{a,b} (\theta,\alpha)=
t^{(i)}_{a,a_1}(\theta)\bar{t} ^{(i+1)}_{a_1,b}
(\theta+\alpha) \qquad i \quad \hbox{odd}.
}
The monodromy matrix that correspond to this model
\eqn\edxi{
T^{\rm{(alt)}}_{a,b} (\theta,\alpha)=\hat{
t}^{(1,2)}_{a,a_1}(\theta,\alpha) \hat{t} ^{(3,4)}_{a_1,a_2}
(\theta, \alpha) \ldots
\bar{t} ^{(2N-1,2N)}_{a_{2N-1,b}}
(\theta,\alpha).
}
This operator can be written in the auxiliary space as a matrix
\eqn\edxii{
T^{alt}(\theta, \alpha)=
\pmatrix{
A(\theta, \alpha) & B_2 (\theta, \alpha) & B_3 (\theta, \alpha) \cr
C_2 (\theta, \alpha)& D_{2,2}(\theta, \alpha) & D_{2,3}(\theta,
\alpha) \cr
C_3 (\theta, \alpha)& D_{3,2}(\theta, \alpha) & D_{3,3}(\theta,
\alpha) \cr
},
}
whose elements are operators in the tensorial product of the site
spaces,
\eqn\edxiii{
S=\bigotimes_{i=\hbox{odd}}{s_{i,i+1}},
}
$s_{i,i+1}$ being the tensorial product of site spaces $(i)$ and
$(i+1)$ and isomorphic to the $\{3\}$ and $\{3^*\}$ representation
product .
\eqn\edxiv{
s_{i,i+1}={s_i} \otimes s_{i+1} \sim \{3\} \otimes \{3^*\}.
}

The YBE for  $T^{\rm{ (alt)}}$ can be written in terms of its components
\eqna\edxv{
$$\eqalignno{
&B(\theta) \otimes B(\theta')= {R}^{(2)} (\theta-\theta') \cdot\bigl(
B(\theta')\otimes B(\theta) \bigr) = \bigl( B(\theta')\otimes
B(\theta) \bigr)\cdot  {R}^{(2)} (\theta-\theta'), & \edxv a \cr
&A(\theta) B(\theta')=g(\theta'-\theta) B(\theta')  A(\theta)
-B(\theta) A(\theta') \cdot {\tilde{r}}^{(2)}(\theta'-\theta),& \edxv
b \cr
&D(\theta) \otimes B(\theta')=g(\theta-\theta') (B(v) \otimes
D(\theta)) \cdot{R}^{(2)} (\theta-\theta') - B(\theta) \otimes
({r}^{(2)}(\theta-\theta') \cdot D(\theta') ), &\cr & &\edxv c \cr
}
$$}
where
\eqn\edxvi{
{R}^{(2)}(\theta) =
\left (\matrix{
1 & 0 & 0 & 0 \cr
0 & {d \over a} & {b \over a} & 0 \cr
0 & {b \over a} & {c \over a} & 0 \cr
0 & 0 & 0 & 1 \cr
}\right ) ,\qquad
{r}^{(2)}(\theta)=\pmatrix{
{h}_{-}& 0 \cr
0 & {h}_{+}\cr
},\qquad
{\tilde{r}}^{(2)}(\theta)=
\pmatrix{
{h}_{+} & 0 \cr
0 & {h}_{-} \cr
},
}
and
\eqn\edxvii{
g(\theta)={ a(\theta)  \over b(\theta) },\qquad {h}_{+}(\theta)=
{c(\theta) \over b(\theta)}, \qquad {h}_{-}(\theta)= {d(\theta) \over
b(\theta)}.
}

For the site states, we use the notation
\eqn\edxviii{
u=\left (\matrix{
1\cr
0\cr
0\cr
}\right ),\quad
d=\left (\matrix{
0\cr
1\cr
0\cr
}\right ),\quad
s=\left (\matrix{
0\cr
0\cr
1\cr
}\right ),\quad
\bar{u}=\left (\matrix{
1\cr
0\cr
0\cr
}\right ),\quad
\bar{d}=\left (\matrix{
0\cr
1\cr
0\cr
}\right ),\quad
\bar{s}=\left (\matrix{
0\cr
0\cr
1\cr
}\right ).
}
In order to find the eigenvectors and eigenvalues of
\eqn\edxix{
\tau^{\rm{(alt)}}(\theta)= A(\theta)+ D_{2,2}(\theta)+ D_{3,3}(\theta)
}
we find inspiration us in the NBA method and look for an eigenstate of $A$
that serves as a pseudovacuum. For this purpose, we build the subspace of
$s_{(i, i+1)}$ generated by the  vectors $\mid u, \bar{s} >$ and
$\mid u, \bar{d} >$, that we call $w_i$, and then the subspace
\eqn\edxx{
\Omega = w_1 \otimes w_3\otimes \cdots  \otimes w_{N}
}
of the total space of states of a chain with $2N$ sites.

In a non-homogeneous chain, we have not a state $\parallel v>$ such
that
\eqn\edxixbis{
 D_{i, j}\parallel v>\propto\delta_{i,j} \parallel v>.
}
\noindent For this
reason, the NBA method cannot be used. Our method, instead,  starts
with a state $\parallel 1> \in  \Omega$ verifying
\eqna\edxxi{
$$\eqalignno{
&A(\theta) \parallel 1> = [ a(\theta)]^{N_3}
[\bar{b}(\theta)]^{N^*_3}  \parallel 1>,& \edxxi a \cr
&B_{i}\parallel 1> \neq 0,\qquad i = 2, 3 , &\edxxi b \cr
&C_{i}\parallel 1> = 0,\qquad i = 2, 3, &\edxxi c \cr
&D_{i, j}\parallel 1> \in  \Omega,\qquad i, j= 2, 3, &\edxxi d \cr
}$$
}
$N_3$ $(N^*_3)$  being the number of sites in the representation
$\{3\}$ $(\{3^*\})$.
In order to simplify the exposition of our method, we take $N_3
=N^*_3= N$.

Following the steps inspired in the NBA, we apply r-times the $B$
operators to $\parallel 1>$ and build the state
\eqn\edxxii{
\Psi (\vec{\mu}) \equiv \Psi (\mu_1, \cdots,
\mu_r)=B_{i_1}(\mu_1) \cdots  B_{i_r}(\mu_r)
X_{i_1,\cdots,i_r} \parallel 1>\equiv B(\mu_1) \otimes \ldots \otimes
B(\mu_r) X \parallel 1>,
}
$X_{i_1,\cdots,i_r} $ being a $r$-tensor that, together with the
values of the spectral parameters $\mu_1 ,\cdots, \mu_r$, will be
determined at the end.

The action  of $A(\mu)$ and $D_{i,i}(\mu)$ on $\Psi$ is found by
pushing them to the right  through the $B_{i_j}(\mu_j)$'s using the
commutations rules \edxv {b,c}. Two types of terms arise  when $A$
and $D_{i,j}$ pass through $B$'s: the wanted and unwanted terms,
similar to obtained in the NBA method. The first one comes from
the first terms of \edxv {b,c}. In this type of terms the $A$ or
$D_{i,i}$ and the $B$'s keep their original arguments and give a
state proportional to $\Psi$. The terms coming from the second terms
in \edxv{b,c} are called unwanted since they contain $B_i(\mu)$ and
so they never give a state proportional to $\Psi$; so, they must
cancel each other out when we sum the trace of $T^{alt}$.
The wanted term obtained by application of $A$ is
\eqn\edxxiii{
[a(\mu)]^{N_3}
[\bar{b}(\mu)]^{N^*_3}\prod_{j=1}^{r}{g(\mu_j-\mu)}B_{i_1}(\mu_1)
\cdots B_{i_r}(\mu_r) X_{i_1,\cdots,i_r} \parallel 1>,
}
and the $k$-th unwanted term
\eqn\edxxiv{
\eqalign{
-[a(\mu_k)]^{N_3} [\bar{b}(\mu_k)]^{N^*_3}
\prod_{j=1 \atop j\neq k}^{r}{g({\mu}_{j} - {\mu}_{k})} &
\left( B(u) \tilde{r}^{(2)}({\mu}_{k} - u)  \right)
\otimes {B}({\mu}_{k+1})\otimes \cdots \cr
&\cdots\otimes {B}({\mu}_{r})\otimes {B}({\mu}_{1})   \otimes
{B}({\mu}_{k-1}) {M}^{(k-1)} X \parallel 1>, \cr
}}
$M$ being the operator arising by repeated application of \edxv {a}\ ,
\eqn\exxv{
{B}({\mu}_{1})\otimes \cdots \otimes
{B}({\mu}_{r})={B}({\mu}_{k+1})\otimes \cdots
{B}({\mu}_{r})\otimes {B}({\mu}_{1}) \cdots  \otimes {B}({\mu}_{k-1})
{M}^{(k-1)}.
}

The application of the operators $D_{i,j}(\mu)$ to the state $\Psi
(\vec{\mu})$ is a little more laborious but straightforward. The
wanted term results
\eqn\edxxvi{
\eqalign{
\bigl[
&D_{k, j}(\mu)B_{i_1}(\mu_1) \cdots  B_{i_r}(\mu_r)
X_{i_1,\cdots,i_r} \parallel 1> \bigr]_{wanted}=
\prod_{i=1}^{r}{g(\mu-\mu_i)} B_{j_1}(\mu_1) \cdots \cr
&\cdots  B_{j_r}(\mu_r) {R}_{ j_r, a_r}^{(2) a_{r-1}, i_r} (\mu-\mu_r)
 \cdots {R}_{  j_2, a_2}^{(2) a_{1}, i_2} (\mu-\mu_2) \cdot
{R}_{ j_1, a_1}^{(2) j, i_1} (\mu-\mu_1)
D_{k, a_r} X_{i_1,\cdots,i_r} \parallel 1>, \cr
}}
where the $R^{(2)}$'s product is taken in the auxiliary space and has
the form
\eqn\edxxvii{
\Phi(\mu, \vec{\mu})_{a_r, j}\equiv{R}_{ j_r, a_r}^{(2) a_{r-1},
i_r}  \cdots
{R}_{  j_2, a_2}^{(2) a_{1}, i_2}  \cdot
{R}_{ j_1, a_1}^{(2) j, i_1} =
\pmatrix{
\alpha(\mu, \vec{\mu}) & \beta(\mu, \vec{\mu}) \cr
\gamma(\mu, \vec{\mu}) & \delta(\mu, \vec{\mu}) \cr
}.
}

The action of  $D_{k,j}$ with $k\neq j$ on $\parallel 1>$ is not
zero. This is the main difference with the models that can be solved
by NBA. Then, we try to diagonalize the matrix product
\eqn\edxxviii{
F(\mu, \vec{\mu})=D(\mu) \cdot \Phi(\mu, \vec{\mu})=
\pmatrix{
A^{(2)}(\mu, \vec{\mu}) & B^{(2)}(\mu, \vec{\mu}) \cr
C^{(2)}(\mu, \vec{\mu}) & D^{(2)}(\mu, \vec{\mu}) \cr
}.
}
By taking the terms in \edxxvi\ with $k=j$ and adding them for $k=2$
and $3$, we obtain the wanted term
\eqn\edxxix{
\prod_{j=1}^{r}{g(\mu-\mu_j)}B_{i_1}(\mu_1) \cdots
B_{i_r}(\mu_r) \tau_{(2)}(\mu, \vec{\mu})X_{i_1,\cdots,i_r} \parallel
1>,
}
where
\eqn\edxxx{
\tau_{(2)}(\mu, \vec{\mu})=\trace(F)=A^{(2)}(\mu,
\vec{\mu})+D^{(2)}(\mu, \vec{\mu}).
}

In the same form, the $k$-th unwanted term results
\eqn\edxxxi{
\eqalign{
-\prod_{j=1 \atop j\neq k}^{r}{g({\mu}_{k} - {\mu}_{j})} &
\left( B(\mu) {r}^{(2)}(\mu-{\mu}_{k} )  \right)
\otimes {B}({\mu}_{k+1})\otimes \cdots \cr
&\cdots\otimes {B}({\mu}_{r})\otimes {B}({\mu}_{1})   \otimes
{B}({\mu}_{k-1}) {M}^{(k-1)}   \tau_{(2)}(\mu_k, \vec{\mu}) X
\parallel 1>. \cr
}}
The  sum of the wanted terms and the cancellation of the unwanted
terms give us the relations
\eqn\edxxxii{
 \tau_{(2)}(\mu, \vec{\mu}) X \parallel 1> =  \Lambda_{(2)}(\mu,
\vec{\mu}) X \parallel 1>
}
and
\eqn\edxxxiii{
 \Lambda_{(2)}(\mu_k, \vec{\mu})=
[a(\mu_k)]^{N_3} [\bar{b}(\mu_k)]^{N^*_3}
\prod_{j=1 \atop j\neq k}^{r}{{g({\mu}_{j} - {\mu}_{k}) \over
g({\mu}_{k} - {\mu}_{j})}}.
}

\noindent We must now diagonalize \edxxxii.

The state $ \parallel 1> \in \Omega$ and the tensor
$X_{i_1,\cdots,i_r},  (i_j =2,3)$ lies in a space with $2^r$
dimensions, tensorial product of $r$ two-dimensional spaces
$C_l, l=1\cdots r$, generated  by the vectors
\eqn\edxxxiv{
e_l^1=\left (\matrix{
1\cr
0\cr
}\right )_l , \qquad
e_l^2=\left (\matrix{
0\cr
1\cr
}\right )_l , \qquad  l=1\cdots r.
}
Then, the vector $X \parallel 1>$ yields in a space $\Omega^{(2)}$
with $2^{r+N}$ dimensions. In this space, we take the element
\eqn\edxxxv{
\parallel 1>^{(2)}=e_1^1\otimes e_2^1
 \cdots
\otimes
e_r^1\otimes
|u \bar{s}>_1\otimes\cdots\otimes|u \bar{s}>_N ,
}
which is annihilated by $C^{(2)}(\mu, \vec{\mu})$. (Note that the
operators $\alpha$, $\beta$, $\gamma$ and $\delta$ of  {\edxxvii} act
on the first part of $\parallel 1>^{(2)}$ and the operators $D_{i,j}$ on
the second part ). The application of the operators $A^{(2)}$ and
$D^{(2)}$ gives
\eqna\edxxxvi{
$$\eqalignno{
&A^{(2)} (\mu, \vec{\mu})\parallel 1>^{(2)}=[b(\mu)]^{N_3}
[\bar{b}(\mu)]^{N^*_3}\parallel 1>^{(2)}, &\edxxxvi a \cr
&D^{(2)} (\mu, \vec{\mu})\parallel 1>^{(2)}=\prod_{i=1}^{r}{{1 \over
g(\mu-\mu_i)}}
[a(\mu)]^{N_3} [\bar{b}(\mu)]^{N^*_3}\parallel 1>^{(2)},
&\edxxxvi b \cr
}$$}
The important fact is that $F(\mu, \vec{\mu})$ verifies the YBE with
the $R^{(2)}$ matrix given in \edxvi,
\eqn\edxxxvii{
R^{(2)}(\mu-\mu') [F(\mu, \vec{\mu}) \otimes F(\mu', \vec{\mu})]=
 [F(\mu', \vec{\mu}) \otimes F(\mu, \vec{\mu})]R^{(2)}(\mu-\mu'),
}
which, in a second step, permits us to solve the system. From this
equation, we obtain the commutation rules
\eqna\edxxxviii{
$$\eqalignno{
&A^{(2)} (\mu) \cdot B^{(2)} (\mu')=g(\mu'-\mu) B^{(2)} (\mu') \cdot
A^{(2)} (\mu )-
h_+(\mu'-\mu) B^{(2)} (\mu ) \cdot A^{(2)} (\mu') ,&\edxxxviii a \cr
&D^{(2)} (\mu ) \cdot B^{(2)} (\mu')=g(\mu-\mu') B^{(2)} (\mu') \cdot
D^{(2)} (\mu )-
h_+(\mu-\mu') B^{(2)} (\mu) \cdot D^{(2)} (\mu') .&\edxxxviii b \cr
}$$}

In this second step, we build the vector
\eqn\edxxxix{
\Psi^{(2)}(\vec{\lambda},\vec{\mu})=B^{(2)} (\lambda_1, \vec{\mu})
\cdots
B^{(2)} (\lambda_s, \vec{\mu})\parallel 1>^{(2)} .
}
The action of $A^{(2)} (\lambda, \vec{\mu})$ on $\Psi^{(2)}$ gives
the wanted term
\eqn\edxxxx{
[b(\lambda)]^{N_3} [\bar{b}(\lambda)]^{N^*_3}
\prod_{i=1}^{s}{g(\lambda_i - \lambda})  B^{(2)}(\lambda_1,\vec{\mu})
\ldots B^{(2)}(\lambda_s,\vec{\mu}) \parallel 1>^{(2)},
}
and the $k$-th unwanted term
\eqn\edxxxxi{
\eqalign{
-h_{+}(\lambda_k-\lambda ) [b(\lambda_k)]^{N_3}
[\bar{b}(\lambda_k)]^{N^*_3} & \prod_{i=1\atop i\neq k}^{s}
{g(\lambda_i - \lambda_k)}  B^{(2)}(\lambda,\vec{\mu})
B^{(2)}(\lambda_{k+1},\vec{\mu}) \ldots    \cr &\ldots
B^{(2)}(\lambda_{k-1},\vec{\mu}) \parallel 1>^{(2)}. \cr
}}
In the same form, the action of $D^{(2)} (\lambda, \vec{\mu})$ on
$\Psi^{(2)}$ gives the wanted term
\eqn\edxxxxii{
[b(\lambda)]^{N_3} [\bar{a}(\lambda)]^{N^*_3}
\prod_{i=1}^{s}{g( \lambda}-\lambda_i)
\prod_{j=1}^{r}{1  \over g(\lambda-\mu_j)}
B^{(2)}(\lambda_1,\vec{\mu}) \ldots B^{(2)}(\lambda_s,\vec{\mu})
\parallel 1>^{(2)},
}
and the $k$-th unwanted term
\eqn\edxxxxiii{
\eqalign{
-h_{-}(\lambda-\lambda_k) [b(\lambda_k)]^{N_3}
[\bar{a}(\lambda_k)]^{N^*_3} & \prod_{i=1\atop i\neq
k}^{s}{g(\lambda_k - \lambda_i)}
\prod_{j=1}^{r}{1  \over g(\lambda_k-\mu_j)}
B^{(2)}(\lambda,\vec{\mu})
B^{(2)}(\lambda_{k+1},\vec{\mu}) \ldots    \cr &\ldots
B^{(2)}(\lambda_{k-1},\vec{\mu}) \parallel 1>^{(2)}. \cr
}}
The cancellation of the unwanted terms and the sum of the wanted
terms give us the equations
\eqn\edxxxxiv{
\biggl[ { {\bar{a}(\lambda_k)} \over {\bar{b}(\lambda_k)} } \biggr]
^{N^*_3}
\prod_{j=1}^{r}{1  \over g(\lambda_k-\mu_j)}=
\prod_{i=1\atop i\neq k}^{s}{ {g(\lambda_i - \lambda_k)} \over
{g(\lambda_k - \lambda_i)}}, \qquad k=1,\ldots,s,
}
and
\eqn\edxxxxv{
\Lambda_{(2)}(\mu_k, \vec{\mu})=
\prod_{i=1}^{s}{ g(\lambda_i-\mu_k)}
[b(\mu_k)]^{N_3} [\bar{a}(\mu_k)]^{N^*_3}.
}

Then, by comparing equations \edxxxiii\ and \edxxxxv\  and calling
$\bar{g}(\theta)={\bar{a}(\theta) / \bar{b}(\theta)}$, we obtain the
coupled Bethe equations
\eqna\edxxxxvi{
$$\eqalignno{
&[\bar{g}(\lambda_k)]^{N^*_3} =\prod_{j=1}^{r}{g(\lambda_k-\mu_j)}
\prod_{i=1\atop i\neq k}^{s}{ {g(\lambda_i - \lambda_k)} \over
{g(\lambda_k - \lambda_i)}}, &\edxxxxvi a \cr
&[g(\mu_l)]^{N_3} =
\prod_{j=1\atop j \neq l}^{r}{ {g(\mu_l -\mu_j)} \over
{g(\mu_j-\mu_l)}}
\prod_{i=1}^{s}{g(\lambda_i-\mu_l)}, &\edxxxxvi b \cr
}$$}
and the eigenvalue of the trace of  $T^{\rm{(alt)}}$
\eqn\edxxxxvii{
\eqalign{
\Lambda(\mu)&=
[a(\mu)]^{N_3} [\bar{b}(\mu)]^{N^*_3}\prod_{j=1}^{r}{g(\mu_j-\mu)}+
\cr
&[b(\mu)]^{N_3}  \prod_{j=1}^{r}{g(\mu-\mu_j)}
\biggl[ [\bar{b}(\mu)]^{N^*_3} \prod_{i=1}^{s}{g(\lambda_i-\mu)} +
[a(\mu)]^{N_3} \prod_{i=1}^{s}{g(\mu-\lambda_i)}
\prod_{j=1}^{r}{{1 \over g(\mu-\mu_j)}} \biggr],  \cr
}}
that is the solution to the spectrum of our problem.

The hamiltonian of the alternating chain can be obtained with \ebxvi\ and
\ebxvii\ . The results for $h^{(1)}$ and $h^{(2)}$ are
\eqn\ehi{
h^{(1)}_{i, i+1} = {{\sinh \gamma (1+2 \cosh \gamma)}
\over {2 (\cosh (3\gamma)-1)}}
\sum_{\alpha =1}^{8} J_{\alpha} \lambda_{i}^{\alpha} \otimes
	\bar{\lambda}_{i+1}^{\alpha}
}
and
\eqnn\ehii{
$$\eqalignno{h_{i,i+1,i+2}^{(2)} =& \sum_{\alpha=1}^{8} m_{\alpha}
	 I_{i} \otimes \bar{\lambda}_{i+1}^{\alpha} \otimes \lambda_{i+2}^{\alpha}
	 +\sum_{\alpha=1}^{8} m'_{\alpha}
	 \lambda_{i}^{\alpha} \otimes I_{i+1} \otimes \lambda_{i+2}^{\alpha}
  \cr
&+ k \left( \lambda_{i}^{3} \otimes I_{i+1} \otimes
	 \lambda_{i+2}^{8} - \lambda_{i}^{8} \otimes I_{i+1} \otimes
	 \lambda_{i+2}^{3} \right) + k' f_{i,i+1,i+2} \, , &\ehii \cr
}$$}
where we have used the Gell-Mann matrices $\lambda$ and $\bar{\lambda}$ for
the $ \{3\}$ and  $ \{3^*\}$ representations respectively, being the
coefficients,
\eqna\ehiii{
$$\eqalignno{m_{\alpha} =& \left\{
\matrix{{{\sinh \gamma (1+2\cosh \gamma)} \over
{2 (\cosh (3\gamma)-1)}} & {\rm{if}} &
	 	\alpha \neq 3,8, \cr
	 	{{\sinh \gamma (-1+4\cosh^{2} \gamma)} \over {2 (\cosh (3\gamma)-1)}} &
	 	{\rm{if}} &  \alpha=3,8, \cr} \right. &\ehiii a\cr
m'_{\alpha} =& \left\{
\matrix{{{\sinh^{2} {\gamma \over 2} (1+2\cosh \gamma) (3+2 \cosh
	 	\gamma)} \over {2 \sinh \gamma (\cosh (3\gamma)-1)}} &{\rm{if}}&
	 	\alpha \neq 3,8,  \cr
	 	{{\sinh^{2} {\gamma \over 2} (1+2\cosh \gamma) (3+\cosh
	 	\gamma +\cosh (2\gamma))} \over {2 \sinh \gamma (\cosh
	 	(3\gamma)-1)}} &
	 	{\rm{if}} &  \alpha=3,8, \cr} \right. &\ehiii b\cr
k =& {{\sqrt{3} \sinh^{2} {\gamma \over 2} (1+2\cosh
	 \gamma)^{2}}\over {4 (\cosh (3\gamma)-1)}},	 &\ehiii c\cr
k' =& {{3 \sinh^{2} {\gamma \over 2} (1+2\cosh
	 \gamma)} \over {\sinh \gamma (\cosh (3\gamma)-1)}}.	&\ehiii d\cr
}$$}
The term $f_{i,i+1,i+2}$ is
\eqnn\ehiv{
$$\eqalignno{f_{i,i+1,i+2}  =& \sum_{\mu,\nu,\rho =1}^{8}
	 d_{\mu,\nu,\rho}(\cosh^{2}({\gamma \over 2}) - {{\sinh \gamma}
	 \over {4}}
	 \epsilon_{\mu,\nu,\rho}) \lambda_{i}^{\mu} \otimes
	 \bar{\lambda}_{i+1}^{\nu} \otimes \lambda_{i+2}^{\rho} \cr
    & + \sum_{\alpha =1}^{8} \left\{ w_{3,\alpha}
   \left( \lambda_{i}^{3} \otimes
	 \bar{\lambda}_{i+1}^{\alpha} \otimes \lambda_{i+2}^{\alpha} -
	  \lambda_{i}^{\alpha} \otimes
	 \bar{\lambda}_{i+1}^{\alpha} \otimes \lambda_{i+2}^{3} \right)
	 \right.  \cr
	 & + w_{8,\alpha} \left( \lambda_{i}^{8} \otimes
	 \bar{\lambda}_{i+1}^{\alpha} \otimes \lambda_{i+2}^{\alpha} -
	  \lambda_{i}^{\alpha} \otimes
	 \bar{\lambda}_{i+1}^{\alpha} \otimes \lambda_{i+2}^{8} \right) \cr
	 & \left. + v_{3,\alpha} \lambda_{i}^{\alpha} \otimes
	 \bar{\lambda}_{i+1}^{3} \otimes \lambda_{i+2}^{\alpha} +
	 v_{8,\alpha} \lambda_{i}^{\alpha} \otimes
	 \bar{\lambda}_{i+1}^{8} \otimes \lambda_{i+2}^{\alpha} \right\}
  + z \left(
	\lambda_{i}^{3} \otimes
	\bar{\lambda}_{i+1}^{8} \otimes \lambda_{i+2}^{3} \right. \cr
	& \left. + \lambda_{i}^{3} \otimes
	\bar{\lambda}_{i+1}^{3} \otimes \lambda_{i+2}^{8} +
	\lambda_{i}^{8} \otimes
	\bar{\lambda}_{i+1}^{3} \otimes \lambda_{i+2}^{3} -
	\lambda_{i}^{8} \otimes
	\bar{\lambda}_{i+1}^{8} \otimes \lambda_{i+2}^{8}
	 \right), &\ehiv \cr
}$$}
where
\eqna\ehv{
$$\eqalignno{\vec{w}_{3} =& {{\sinh \gamma}\over{4}}
\pmatrix{2 & 2 & 0 & -1 & -1 & -1 & -1 & 0 \cr}  ,  &\ehv a\cr
  \vec{w}_{8} =& {{\sqrt{3}\sinh \gamma}\over{4}}
  \pmatrix{0 & 0 & 0 & -1 & -1 & 1 & 1 & 0 \cr}    ,&\ehv b\cr
 \vec{v}_{3} = &{{-\sinh^{2}{\gamma \over 2}}\over{2}}
 \pmatrix{0 & 0 & 0 & -1 & -1 & 1 & 1 & 0 \cr}  , &\ehv c\cr
 \vec{v}_{8} =&{{\sinh^{2}{\gamma \over 2}} \over {2\sqrt{3}}}
 \pmatrix{2 & 2 & 0 & -1 & -1 & -1 & -1 & 0 \cr}  ,&\ehv d\cr
 z =& {{\sinh^{2} \gamma}\over {\sqrt{3}}} ,  &\ehv e\cr
}$$}
$d_{\mu,\nu,\rho}$ are the totally symmetric structure constants
of $SU(3)$, and $\epsilon_{\mu,\nu,\rho}$ is the totally antisymmetric
tensor.

%
%
%
%
%
%
%

As a first generalization, we can apply now the the method to a chain
that mixes the  $(1,0)\equiv \{3\}$ and $(m_1, m_2)$ representations.
In this model we take again the $(1, 0)$ representation as auxiliary
space; then we have the same $R$-matrix \edviii\ for the YBE.

The highest weight of the $(m_1, m_2)$ representation is
\eqn\edxxxxviii{
\Lambda_{h}={{2 m_1 + m_2} \over 3} \alpha_1 +{{m_1 +2 m_2} \over 3}
\alpha_2  ,
}
where $\alpha_1$ and $\alpha_2 $ are the simple roots of $su(3)$.

Through \eciii{a\hbox{--} c}, \eciv, together with the commutation rules of
$su(3)$, it
is possible to know the action of elements of the $\bar{t}$ matrix
on the highest weight vector. We obtain
\eqna\edxxxxix
$$\eqalignno{{\bar{t}}_{1,1}(\theta)| \Lambda_{h} >&=
\bar{a}(\theta)  |\Lambda_{h} >, &\edxxxxix a  \cr
{\bar{t}}_{2,2}(\theta)| \Lambda_{h} >&= \bar{b_1}(\theta) |
\Lambda_{h} >, &\edxxxxix b \cr
{\bar{t}}_{3,3}(\theta)|\Lambda_{h} >&= \bar{b_2}(\theta) |
\Lambda_{h} >, &\edxxxxix  c \cr}
$$
where
\eqna\edxxxxx
$$\eqalignno{
\bar{a}(\theta) &=\sinh(
{3 \over 2}\theta+({2 \over 3}m_{1}+{1 \over 3}m_{2}+
{1 \over 3})\gamma) ,&\edxxxxx a  \cr
\bar{b_1}(\theta) &=\sinh
({3 \over 2}\theta+(-{1 \over 3}m_{1}+{1 \over 3}m_{2}+
{1 \over 3})\gamma), &\edxxxxx b  \cr
\bar{b_2}(\theta) &=\sinh
({3 \over 2}\theta+(-{1 \over 3}m_{1}-{2 \over 3}m_{2}+
{1 \over 3})\gamma). &\edxxxxx c  \cr}
$$

As before, we group neighbor sites and build the monodromy
operator $T$ that can be represented by a matrix in the auxiliary
space as in \edxii . The two sites space is now
\eqn\edxxxxxiii{
s_{i, i+1} \sim (1, 0)\otimes(m_1, m_2).
}
In this space, the subspace $w_i$ is now generated by the highest
weight of the $(1,0)$ representation and the subspace $V$ generated by the
states

\eqn\edbis{
 \{|\Lambda_{h} >, f_{2} |\Lambda_{h} >,
f_{2}^{2} |\Lambda_{h} >, \ldots \},
}
where $f_{2}$ is the generator of $sl(3)$ in the $(m_{1},m_{2})$
representation.

We form the subspace $\Omega$ as in \edxx\ and
built the state  $\parallel 1> \in \Omega$ which must satisfy
\eqna\edxxxxxiv{
$$\eqalignno{
&A(\theta) \parallel 1> \propto  \parallel 1>,& \edxxxxxiv a \cr
&D_{i,i}(\theta) \parallel 1> \propto  \parallel 1>,\qquad i= 2, 3,&
\edxxxxxiv b \cr
&B_{i}\parallel 1> \neq 0, \qquad i = 2, 3 , &\edxxxxxiv c \cr
&C_{i}\parallel 1> = 0,\qquad i = 2, 3, &\edxxxxxiv d \cr
&D_{i, j}\parallel 1> \in  \Omega, \qquad i, j= 2, 3,\quad i \neq j
.&\edxxxxxiv e
\cr
}$$
}
Then, the states $\Psi (\vec{\mu})$ analogous to \edxxii\ are
\eqn\edxxxxxv{
\Psi (\vec{\mu}) \equiv \Psi (\mu_1, \cdots,
\mu_r)=B_{i_1}(\mu_1) \cdots  B_{i_r}(\mu_r)
X_{i_1,\cdots,i_r} \parallel 1>\equiv B(\mu_1) \otimes \ldots \otimes
B(\mu_r) X \parallel 1>.
}

As the YBE depends on the $R$ matrix, we have for
the new monodromy matrix the same commutations rules \edxv{a\hbox{--}c}\ as
before; then we can repeat the same steps, the only difference being in
the action of the operators of the monodromy matrix on the state $ \parallel
1>$. The BEs that we obtain in this case are
\eqna\edxxxxxvi{
$$\eqalignno{&[g(\mu_{k})]^{N}[\bar{g}_{1}(\mu_{k})]^{N}
=\prod_{j=1 \atop j \neq
k}^{r} {g(\mu_{k}-\mu_{j}) \over g(\mu_{j}-\mu_{k})}\prod_{i=1}^
{s}g(\lambda_{i}-\mu_{k}),&\edxxxxxvi a \cr
&[\bar{g}_{2}(\lambda_{k})]^{N}=\prod_{j=1}^{r}g(\lambda_{k}-\mu_{j})
\prod_{i=1 \atop i \neq
k}^{s} {g(\lambda_{i}-\lambda_{k}) \over g(\lambda_{k}-\lambda_{i})},
&\edxxxxxvi b
\cr}
$$}
where $\mu_i$, $i=1,\cdots, r$, and $\lambda_j$, $j=1,\cdots, s$, are
the roots of
the ansatz, the function $g$ is given in \edxvii , and
\eqna\edxxxxxvii {
$$\eqalignno{
&\bar{g}_1(\theta)={ {  \bar{a}(\theta)  }\over  \bar{b}_{1}(\theta)},
&\edxxxxxvii  a \cr
&\bar{g}_2(\theta)={ {  \bar{b}_{2}(\theta)  }\over
\bar{b}_{1}(\theta)}.
&\edxxxxxvii  b \cr}
$$}

The procedure can be generalized to chains that mix non-fundamental
representations $(m_1, m_2)$ and $(m'_1,m'_2)$, irrespective of the
number of sites and their distribution
in the representations. For this purpose, it is necessary to build
the monodromy matrix
following an analogous process to used before. If we use a dashed line
for the
representation $({m'}_1, {m'}_2)$, the monodromy matrix $T^{\rm
gen}(\theta)$
can be represented  graphically as shown in figure 7.
\midinsert
\bigskip
\centerline{\epsfxsize=10cm  \epsfbox{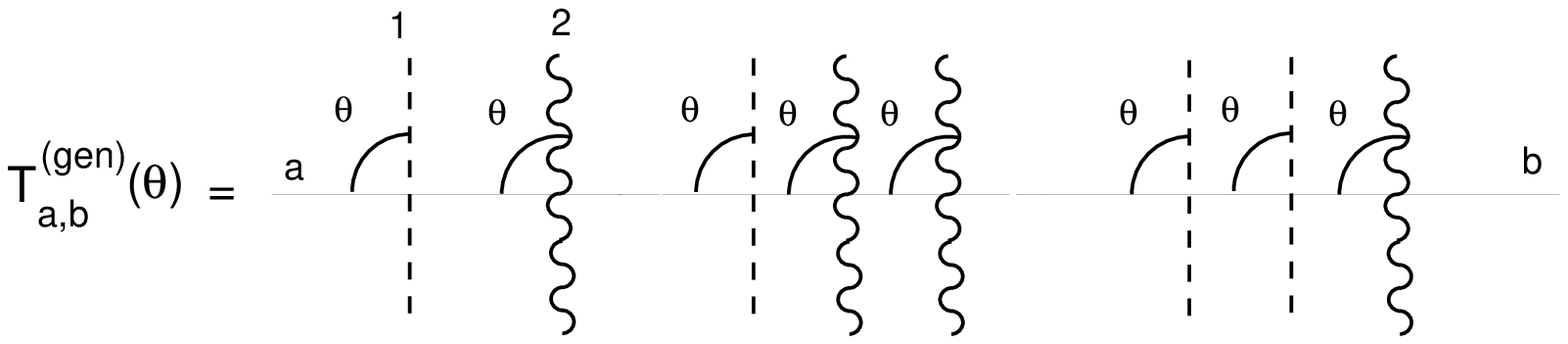}}
\centerline{Figure 7}
\endinsert
\bigskip

By calling $N_1$ and  $N_2$ the number of sites in the
representations $({m}_1,
{m}_2)$ and $({m'}_1, {m'}_2)$ respectively, we find the BE for this
general chain
\eqna\edxxxxxviii
$$\eqalignno{[\widetilde{g}_{1}(\mu_{k})]^{{N}_1}[\bar{g}_{1}(\mu_{k})]^{{N}_2}
&=\prod_{j=1 \atop j \neq
k}^{r} {g(\mu_{k}-\mu_{j}) \over g(\mu_{j}-\mu_{k})}\prod_{i=1}^
{s}g(\lambda_{i}-\mu_{k}) &\edxxxxxviii a \cr
[\widetilde{g}_{2}(\lambda_{k})]^{{N}_1}[\bar{g}_{2}(\lambda_{k})]^{{N}_2}&=
\prod_{j=1}^{r}g(\lambda_{k}-\mu_{j})
\prod_{i=1 \atop i \neq
k}^{s} {g(\lambda_{i}-\lambda_{k}) \over g(\lambda_{k}-\lambda_{i})}
&\edxxxxxviii b \cr}
$$
where $\bar{g}_{1}$ and $\bar{g}_{2}$ are given in \edxxxxxvii{a,b}, and
 $\widetilde{g}_{1}$ and $\widetilde{g}_{2}$ are the same as the
previous ones with
$(m_{1},m_{2})$ replaced by $(m'_{1},m'_{2})$.

In the light of this, the generalization for the case of
mixed chains with more than two different representations seems
simple, although the physical
models that they represent will be less local and the interaction
more complex.

Also we can conjecture about the solution of a non-homogeneous chain combining different representations of $su(n)$, each representation introduces $(n-1)$   functions $g_i$ similar to \edxxxxxvii {a,b}\ (that we call source functions). The BE are obtained applying the MBA with $(n-1)$ steps, then each solution will have a set of $(n-1)$ equations (the same number of dots in its Dynkin diagram). The first member of  the equations will be a product of the respective source functions powered to the number of sites of each representation and the second a product of $g$ functions coming from the YBE similar to \edxxxxxviii {a,b}\ .

%
%
%
%
\newsec{Thermodynamic limits of solutions and analysis of Bethe
equations}

In this section, we are going to discuss the solutions of the
$\{3\}$-$\{3^*\}$ model  given by the equations \edxxxxv\ in the limit
for very large $N$. For that discussion, it is convenient to set the
parametrization of the spectral parameters
\eqna\eei{
$$\eqalignno{
&{3 \over 2}\mu_j= i v_j^{(1)}- {\gamma \over 2},&\eei a \cr
&{3 \over 2}\lambda_j= i v_j^{(2)}- \gamma  ,&\eei b \cr
}
$$}
and $N=N_{3}+N_{3}^{*}$ the length of the chain.

Using such parametrization, the Bethe equations \edxxxxvi{a, b}\ can be written

\eqna\eeii{
$$\eqalignno{
&\Biggl[{\sin{(v_k^{(2)}+ i {\gamma \over 2})} \over \sin{(v_k^{(2)}-
i {\gamma \over 2})}}\Biggr]^{N_3^*}=
-\prod_{j=1}^{r}{{\sin{(v_k^{(2)}-v_j^{(1)} - i {\gamma \over 2})}
\over \sin{(v_k^{(2)}-v_j^{(1)} + i {\gamma \over 2})}}}
\prod_{i=1 \atop i\neq k}^{s}{{\sin{(v_i^{(2)}-v_k^{(2)} - i \gamma
)} \over \sin{(v_i^{(2)}-v_k^{(2)} + i \gamma )}}} ,&\eeii a \cr
&\Biggl[{\sin{(v_k^{(1)}- i {\gamma \over 2})} \over \sin(v_k^{(1)}+
i {\gamma \over 2})}\Biggr]^{N_3}=
-\prod_{j=1 \atop j \neq k}^{r}{{\sin{(v_k^{(1)}-v_j^{(1)} - i \gamma
)} \over \sin{(v_k^{(1)}-v_j^{(1)} + i \gamma )}}}
\prod_{i=1}^{s}{{\sin{(v_i^{(2)}-v_k^{(1)} - i {\gamma \over 2})}
\over \sin{(v_i^{(2)}-v_k^{(1)} + i {\gamma \over 2})}}} ,&\eeii b
\cr
}
$$}
%

In this regime, the roots must be considered in the interval $(-\pi / 2,
\pi /2)$. Then, we define the function
\eqn\eeiii{
\phi (\chi, \alpha)= i \ln{{\sin{(\chi + i \alpha)} \over \sin{(\chi
- i \alpha)}}}
}
and taking logarithms in \eeii{a, b}\ we obtain
\eqna\eeiv{
$$\eqalignno{
&N_3^* \phi (v_k^{(2)}, {\gamma \over 2})+
\sum_{j=1}^{r}{\phi(v_k^{(2)}-v_j^{(1)}, {\gamma \over 2})}
-\sum_{i=1}^{s}{\phi(v_k^{(2)}-v_i^{(2)}, \gamma)} =2 \pi I_k^{(2)} ,
\ 1\leq  k\leq s,       &\eeiv a \cr
&N_3 \phi(v_k^{(1)}, {\gamma \over 2})-
\sum_{j=1}^{r}{\phi(v_k^{(1)}, {\gamma \over 2})}
+\sum_{i=1}^{s}{\phi(v_k^{(2)}-v_i^{(1)}, {\gamma \over 2})} =2 \pi
I_k^{(1)},  \ 1\leq  k \leq r,  &\eeiv b \cr
}
$$}
%
%
%
where $I_k^{(1)}$ and $I_k^{(2)}$ are half-integers.

In the thermodynamic limit $N\rightarrow \infty$, the roots tend to
have continuous distributions. Unlike what happens in other cases, we
cannot distinguish between the roots coming from the different types
of representations, this can be noted by simple inspection of the
equations of the ansatz. Due to that, we define two root densities,
one for each level,
\eqn\eev{
\rho_l(v_{j}^{(l)})=\lim_{N_3\rightarrow \infty}{{1 \over
{N_3(v_{j+1}^{(l)}-v_{j}^{(l)})}}}, \qquad l=1,2,
}

Let it be
\eqna\eevi{
$$\eqalignno{
&Z_{N_3}(v)={1 \over {2 \pi}} \Bigl[
\phi (v, {\gamma \over 2})-
{1 \over N_3} \sum_{j=1}^{r}{\phi(v-v_j^{(1)}, \gamma )}+
{1 \over N_3} \sum_{j=1}^{s}{\phi(v-v_j^{(2)}, {\gamma \over 2})}
\Bigr],
&\eevi a \cr
&Z_{N_3^*}(v)={1 \over {2 \pi}} \Bigl[
\phi (v, {\gamma \over 2})-
{1 \over N_3^*} \sum_{j=1}^{s}{\phi(v-v_j^{(2)}, \gamma )}+
{1 \over N_3^*} \sum_{j=1}^{r}{\phi(v-v_j^{(1)}, {\gamma \over 2})}
\Bigr],
&\eevi b \cr
}$$}
The no-holes hypothesis for the fundamental state establishes
\eqn\eevii{
I_{k-1}^{(i)}-I_{k}^{(i)}=1, \qquad i=1,2, \qquad \hbox{\rm for all} \quad k,
}
that implies
\eqna\eeviii{
$$\eqalignno{
&Z_{N_3}(v_k^{(1)})={I_k^{(1)} \over N_3},
&\eeviii a \cr
&Z_{N_3^*}(v_k^{(2)})={I_k^{(2)} \over N_3^*}.
&\eeviii b \cr
}$$}

In the thermodynamic limit and for the fundamental state, the derivative
of these functions are
\eqna\eeix{
$$\eqalignno{
&\sigma^{(1)}(v)\equiv{d \over dv}{Z_{N_3}(v)}\approx {N \over
{N_3}}\rho_1(v),
&\eeix a \cr
&\sigma^{(2)}(v)\equiv{d \over dv}{Z_{N_3^*}(v)}={ N
\over N_3^*}\rho_2(v).
&\eeix b \cr
}$$}
Using the approximation
\eqn\eex{
\lim_{N_3 \rightarrow \infty}{  {1 \over N_3}
\sum_{j} f(v_j^{(k)})  \simeq
\int_{-{\pi \over 2}}^{{\pi \over 2}}
{d\lambda f(\lambda) \rho_k(\lambda)}
},
}
together with \eeix{a,b}\ and \eevi{a,b}, we obtain the system of
equations
\eqna\eexi{
$$\eqalignno{
&\rho_1(\lambda)={1 \over 2 \pi} \biggl[ {N_3 \over N}\phi'
(\lambda,{\gamma \over 2})-
\int_{{\pi  \over 2}}^{-{\pi  \over 2}}{ \phi' (\lambda-\mu,\gamma )
\rho_1(\mu) d\mu}+
\cr
& \qquad\qquad\qquad\qquad\qquad\qquad\qquad\qquad+\int_{{\pi  \over
2}}^{-{\pi  \over 2}}{ \phi' (\lambda-\mu,{\gamma \over 2})
\rho_2(\mu) d\mu}
\biggr] ,
&\eexi a \cr
&\rho_2(\lambda)={1 \over 2 \pi} \biggl[ {N_3^* \over N}\phi'
(\lambda,{\gamma \over 2})-
\int_{{\pi  \over 2}}^{-{\pi  \over 2}}{ \phi' (\lambda-\mu,\gamma )
\rho_2(\mu) d\mu}+ \cr
& \qquad\qquad\qquad\qquad\qquad\qquad\qquad\qquad
+\int_{{\pi  \over 2}}^{-{\pi  \over 2}}{ \phi' (\lambda-\mu,{\gamma
\over 2}) \rho_1(\mu) d\mu}
\biggr] ,
&\eexi b \cr
}$$}
that can be solved by doing the Fourier transform,
\eqna\eexii{
$$\eqalignno{
&\phi(\lambda,\alpha)= \pi +2 \lambda -i \sum_{m=- \infty \atop m
\neq 0}^{\infty}
{{1 \over m} e^{2 i m\lambda -2 |m| \alpha}},
&\eexii a \cr
&\rho_j(\lambda)=\sum_{m \in Z}
{{1 \over 2 \pi} e^{2 i m \lambda}\hat{\rho}_j(m) }.
&\eexii b \cr
}$$}
Introducing these expressions in the integral equations \eexi{a,b}, we
obtain the densities in the Fourier space
\eqna\eexiii{
$$\eqalignno{
&\hat{\rho}_1(m) = 2 {N_3 \over N}{ \sinh{(2 \gamma |m|)}\over
\sinh{(3 \gamma |m| )}}+
2 {N_3^* \over N}
{ \sinh{(\gamma |m|)}\over \sinh{(3 \gamma |m| )}},
&\eexiii a \cr
&\hat{\rho}_2(m) = 2 {N_3 \over N}{ \sinh{(\gamma |m|)}\over \sinh{(3
\gamma |m|) }}+
2 {N_3^* \over N}
{ \sinh{ (2 \gamma |m|)}\over \sinh{(3 \gamma |m| )}},
&\eexiii b \cr
}$$}
when $m \neq 0$, and
\eqna\eexiiir{
$$\eqalignno{
&\hat{\rho}_1(0) = {{2(2N_3 +N_3^*)}\over 3 N}&\eexiiir a \cr
&\hat{\rho}_2(0) = {{2(2N_3 ^*+N_3)}\over 3 N}&\eexiiir b \cr
}$$}
for $m=0$.
We note that for $N_3^*=0$ we have again the known result for a
homogeneous chain. It is interesting  to notice the complementarity of
the solution for $N_3^*=0$ and the solution for $N_3=0$,
\eqna\eexiiirr{
$$\eqalignno{
&\hat{\rho}_1(m)\mid_{N_3=0} =
\hat{\rho}_2(m)\mid_{N_3^*=0},&\eexiiirr a \cr
&\hat{\rho}_2(m)\mid_{N_3=0} =
\hat{\rho}_1(m)\mid_{N_3^*=0}.&\eexiiirr b \cr
}$$}

In the case $N_3=N_3^*=N/2$, that corresponds to our alternating chain,
the densities are given by
\eqn\eexiv{
\hat{\rho}_1(m) =\hat{\rho}_2(m) =
{\cosh{{1 \over 2} m \gamma} \over \cosh{{3 \over 2} m \gamma}}.
}

The free energy is defined by the expression
\eqn\eexv{
\lim_{N\rightarrow \infty}{}f(\theta, \gamma) = -{1 \over N}
\lg{\Lambda(\theta)}.
}
Then, taking the dominant term in $\Lambda(\theta)$,
\eqn\eexvi{
\Lambda_+(\theta)= [ a(\theta)]^{N_3} [
\bar{b}(\theta)]^{N_3^*}\prod_{j=1}^{r}
{g(\mu_j-\theta)},
}
the energy is given in this limit by
\eqn\eexvii{
f(\theta, \gamma)= - {N_3 \over N}\lg{(a(\theta))}- {N_3^* \over N}
\lg{(\bar{b}(\theta))}+
{i \over N}\sum_{j=1}^{r}{\Phi(v_j^{(1)}+i{3 \over 2}\theta,{\gamma
\over 2})}.
}

Doing the change of variable $u=3 \theta /2$ and using equations
\eex\ and \eexiii{a,b}, the free energy can be written in the more
transparent form
\eqna\eexviii{
$$\eqalignno{
f(u,\gamma) =& - { N_3 \over N} \ln (\sinh (u+\gamma)) + {4 \over 3}
{ N_3 \over N}u & \cr
 &+{ 2 N_3 \over N} \sum_{m=1}^{\infty} {e^{-m\gamma} \over m}
	 \sinh (2mu) {\sinh(2m\gamma) \over \sinh(3m\gamma)}& \cr
&- { N_3^* \over N} \ln (\sinh (u+ {3 \over 2}\gamma)) + {2 \over 3}
{ N_3^* \over N}u & \cr
 &+{ 2 N_3^* \over N} \sum_{m=1}^{\infty} {e^{-m\gamma} \over m}
	 \sinh (2mu) {\sinh(m\gamma) \over \sinh(3m\gamma)}.&\eexviii\ \cr
}$$}

As we can see, the free energy is the sum of the individual
contributions of the sites in each representation. So, for $N^*_3=0$
($N_3=0$), we obtain again the results of the homogeneous case in the
representation $\{3\}$ ($\{3^*\}$).

{}From the free energy, we can obtain the energy density in the
fundamental state,
\eqn\eexxix{
{\cal E} =-{d{f} \over d\theta} \bigg|_{\theta=0} = -{3 \over 2}{d{f}
\over du} \bigg|_{u=0}
}
Doing the calculation the result is
\eqna\eexxx{
$$\eqalignno{
 {\cal E} =&-{3 \over 2} \Biggl\{ {N_{3} \over N} \biggl[
	 -\coth \gamma + {4 \over 3} + 4 \sum_{m=1}^{\infty} e^{-m\gamma}
	{\sinh (2m\gamma) \over \sinh (3m\gamma)}
	 \biggr] &  \cr
&  + {N_3^*\over N} \biggl[-\coth ({{3} \over 2}
	 \gamma) + { 2 \over 3} + 4 \sum_{m=1}^{\infty} e^{-m\gamma}
	{ \sinh (m\gamma) \over \sinh (3m\gamma)}
	 \biggr]
   \Biggr\}, &\eexxx\ \cr
}$$}
that again is the sum of the individual contributions of each site
representations.

We can apply the results to the alternating case $(N_3=N_3^*=N/2)$;
the free energy is
\eqna\eexxxi{
$$\eqalignno{
f^{(alt)}(u,\gamma)=&{1 \over 2}  \ln (\sinh (u+\gamma))-{1 \over 2}
 \ln (\sinh (u+{3 \over 2}\gamma)) & \cr
& +\sum_{m=1}^{\infty}{ {e^{-m\gamma}} \over m} \sinh (2mu)
{{\cosh({1 \over 2} m\gamma)} \over \cosh({{3} \over 2} m\gamma)},&
\eexxxi\ \cr
}$$}
and the energy density of the fundamental state
\eqn\eexxxii{
{\cal E}^{(alt)} ={3 \over 4} \biggl(\coth \gamma +
	\coth ({3 \over 2} \gamma)
	\biggl) +{ 3\over 2} -3 \sum_{m=1}^{\infty} e^{-m\gamma}
	 {{\cosh ({1 \over 2} m \gamma)} \over \cosh {3 \over 2}m \gamma)}.
}

The solutions we have given, have been obtained by taking hyperbolic functions
for
the solutions \edii{a\hbox{-}d}\ and \edvii{a\hbox{-}d}\ of the YBE.
By considering the trigonometric solutions of these equations and following the
same steps, we find the BA
equations,
\eqna\eexxxii{
$$\eqalignno{
\biggl[{\sinh(v_{k}^{(1)}-i{\gamma \over 2}) \over
	\sinh(v_{k}^{(1)}+i{\gamma \over 2})}\biggr]^{N_{3}}\!=\!&
	- \prod_{j=1}^{r}{\sinh(v_k^{(1)}-v_j^{(1)}-i\gamma) \over
	\sinh(v_{k}^{(1)}-v_{j}^{(1)}+i\gamma)}
	\prod_{l=1}^{s}{\sinh(v_{l}^{(2)}-v_{k}^{(1)}-i{\gamma \over 2})
\over
	\sinh(v_{l}^{(2)}-v_{k}^{(1)}+i{\gamma \over 2})}
	& \eexxxii a \cr
 \biggl[{\sinh(v_{k}^{(2)}+i{\gamma \over 2}) \over
	\sinh(v_{k}^{(2)}-i{\gamma \over 2})}\biggr]^{N_{3}^*}\!=\!&
	- \prod_{j=1}^{r}{\sinh(v_k^{(2)}-v_j^{(1)}-i\gamma) \over
	\sinh(v_{k}^{(2)}-v_{j}^{(1)}+i\gamma)}
	\prod_{l=1}^{s}{\sinh(v_{l}^{(2)}-v_{k}^{(2)}-i{\gamma \over 2})
\over
	\sinh(v_{l}^{(2)}-v_{k}^{(2)}+i{\gamma \over 2})}
	& \eexxxii b \cr
}
$$}
In this regime, the roots cover all real numbers $(-\infty, \infty)$.
Then, defining an analogous function
\eqn\eexxxiii{
\Phi(x,\alpha) = i\ln {\sinh (x+i\alpha) \over \sinh (x-i\alpha)},
}
we can solve the problem again by using the Fourier transform
\eqna\eexxxiv{
$$\eqalignno{
&\Phi(\lambda,\alpha) = \pi + \int_{-\infty}^{+\infty}
	 {{dk} \over k} \sin (k\lambda) {\sin(k({\pi \over 2}-\alpha)) \over
	 \sin (k{\pi \over 2} )},& \eexxxiv a \cr
& \rho_{j}(\lambda) = {{1} \over 2\pi}\int_{-\infty}^{+\infty}
	 dk e^{ik\lambda} \hat{\rho}_{j}(k) .& \eexxxiv b \cr
}$$}
The no-holes hypothesis for the ground state gives us the densities
\eqna\eexxxv{
$$\eqalignno{
\hat{\rho}_{1}(k) & =  {N_{3} \over N} {\sinh (k\gamma) \over
       \sinh ({ 3\over 2}k\gamma)} + {N_{3}^{*} \over N} {\sinh
	(k {\gamma \over 2})
 \over \sinh ({3 \over 2}k\gamma)},
	& \eexxxv a \cr
	\hat{\rho}_{2}(k) & =  {N_{3} \over N} {\sinh (k{\gamma \over 2})
\over
       \sinh ({ 3\over 2}k\gamma)} + {N_{3}^{*} \over N} {\sinh
	(k \gamma)\over \sinh ({3 \over 2}k\gamma)},
& \eexxxv b \cr
}$$}
and the free energy becomes
\eqnn\eexxxvi{
$$\eqalignno{
 f(u,\gamma)& =  {N_{3} \over N} \biggl\{
	 -\ln \sin (u+\gamma) + 2 \int_{0}^{\infty} {dk \over k}
	{ \sinh(ku) \sinh(k({\pi \over 2}-{\gamma \over 2})) \sinh(k\gamma)
\over
	 \sinh(k{\pi \over 2})\sinh(k{3\gamma \over 2})}
	 \biggr\}   \cr
& +{N_{3}^*\over N} \biggl\{
	 -\ln \sin (u+{ 3\over 2}\gamma) + 2 \int_{0}^{\infty} {dk \over k}
	{ \sinh(ku) \sinh(k({\pi \over 2}-{\gamma \over 2})) \sinh(k\gamma)
\over
	 \sinh(k{\pi \over 2})\sinh(k{3\gamma \over 2})}
	 \biggr\} &\eexxxvi  \cr
}$$}

The density of energy of the ground state is
\eqnn\eexxxvii{
$$\eqalignno{
{\cal E} =& -{ 3\over 2}\biggl\{ {N_{3} \over N}
 \biggl[-\cot \gamma
	 + 2 \int_{0}^{\infty} dk
	{ \sinh(k({\pi \over 2}-{\gamma \over 2})) \sinh(k\gamma) \over
	 \sinh(k{\pi \over 2})\sinh(k{3\gamma \over 2})}
	 \biggr]   \cr
&{N_{3}^* \over N}
 \biggl[-\cot {3 \over 2}\gamma
	 + 2 \int_{0}^{\infty} dk
	{ \sinh(k({\pi \over 2}-{\gamma \over 2})) \sinh(k\gamma) \over
	 \sinh(k{\pi \over 2})\sinh(k{3\gamma \over 2})}
	 \biggr]  \bigg\}&\eexxxvii \cr
}$$}

We can specify these magnitudes for the alternating case
$(N_3=N_3^*=N/2)$; they are
\eqn\eexxxviii{
f^{(alt)}(u,\gamma) =
	 -{1 \over 2}\ln \sin (u+\gamma) -{ 1\over 2}\ln \sin (u+{ 3\over
2}\gamma)
+ \int_{0}^{\infty} {dk \over k}
	{ \sinh(ku) \sinh(k({\pi \over 2}-{\gamma \over 2})) \cosh({k \over
4}\gamma) \over
	 \sinh(k{\pi \over 2})\cosh(k{3\gamma \over 4})}
}
and
\eqn\eexxxix{
{\cal E}^{alt}=-{3 \over 4}\bigr(\cot \gamma -\cot({3 \over 2}
\gamma)\bigr)-
{3 \over 2}
\int_{0}^{\infty} dk
	{ \sinh(k({\pi \over 2}-{\gamma \over 2})) \cosh(k{\gamma \over 4})
\over
	 \sinh(k{\pi \over 2})\cosh(k{3\gamma \over 4})}.
}

We can describe other quantum numbers of the eigenvectors of the transfer
matrix. Let we define the number operators
\eqna\ezi{
$$\eqalignno{\widehat{Y}_{1} =& \widehat{N}_{u}-\widehat{N}_{\bar{u}}, &\ezi
a\cr
 \widehat{ Y}_{2} =& \widehat{N}_{d}-\widehat{N}_{\bar{d}},&\ezi b\cr
}$$}
where
\eqn\ezii{
\widehat{N}_{\alpha} = \sum_{i=1}^{N} 1_{1} \otimes  \ldots \otimes 1\otimes
	(n_{\alpha})_{i} \otimes 1 \otimes \ldots \otimes 1_{N} ,
}
and
\eqn\eziii{
n_{\alpha} | \beta > = \left\{
	\matrix{
		1 & \hbox{\rm if} & \beta = \alpha  \cr
		0 & if & \beta \neq \alpha \cr}
	\right. \,.
}

The operators $\widehat{Y}_{1}$ and $\widehat{Y}_{2}$ commute with the transfer
matrix
\eqn\eziv{
[\widehat{Y}_{i},\tau(\theta)] = 0 \quad i=1,2
 \,.
}

The commutation relations with the $B$-operators are
\eqna\ezv{
$$\eqalignno{[\widehat{Y}_{1},B_{i}(\theta)]=& -B_{i}(\theta), &\ezv a\cr
  [\widehat{Y}_{2},B_{i}(\theta)] =& \delta_{2,i}B_{i}(\theta) \,. &\ezv b\cr
}$$}
Then, if we apply $\widehat{Y}_{1}$ and $\widehat{Y}_{2}$ on the state $\Psi
(\vec{\mu})$, obtained by the aplication of $r$ operators $B$ to the
pseudovacuum state $|| 0 >$ in the first step and $s$ operators in the second
step, we
find
\eqna\ezvi{
$$\eqalignno{\widehat{Y}_{1}\Psi (\vec{\mu})=&(N_{3}-r) \Psi (\vec{\mu}) ,
&\ezvi a\cr
   \widehat{Y}_{2}\Psi (\vec{\mu})=&(r-s) \Psi (\vec{\mu}) \,, &\ezvi b\cr
}$$}
we have the quantum numbers of this problem as
\eqna\ezvii{
$$\eqalignno{N_{u}-N_{\bar{u}}=& N_{3}-r ,&\ezvii a\cr
  N_{d}-N_{\bar{d}}=& r-s  ,&\ezvii b\cr
}$$}
and obviously
\eqn\ezviii{
N_{s}-N_{\bar{s}}= s-N_{3}^{*} ,
}
being $N_{q}$ the eigenvalues of $\widehat{N}_q$, $(q= u, \bar{u},  d, \bar{d},
s, \bar{s})$.

In the thermodynamic limit the fundamental state is characterized by
\eqna\ezviii{
$$\eqalignno{\left({r \over N}\right)_{N \rightarrow \infty}=&
\int_{-A}^{A} \rho_{1}(\lambda) d\lambda = {{2N_{3}+N_{3}^{*}}\over {3N}},
 &\ezviii a\cr
  \left({s \over N}\right)_{N \rightarrow \infty}=&
   \int_{-A}^{A} \rho_{2}(\lambda) d\lambda =
	{{N_{3}+2N_{3}^{*}} \over {3N}},   &\ezviii b\cr
}$$}
and then
\eqna\ezix{
$$\eqalignno{\left({{N_{u} - N_{\bar{u}}}\over{N}}\right)_{N \rightarrow
\infty}=& {{N_{3}-N_{3}^{*}}\over{3N}}, &\ezix a\cr
 \left({{N_{d} - N_{\bar{d}}}\over{N}}\right)_{N \rightarrow
  \infty} =&  {{N_{3}-N_{3}^{*}}\over{3N}}, &\ezix b\cr
 \left({{N_{s} - N_{\bar{s}}}\over{N}}\right)_{N \rightarrow
 \infty} =& {{N_{3}-N_{3}^{*}}\over {3N}} \,. &\ezix c\cr
}$$}
For $N_{3}=0$ or $N_{3}^{*}=0$ we recuperate the no mixing chain
results.

In the alternating chain $(N_{3}=N_{3}^{*}=N/2)$ we obtain
\eqn\ezx{
 \left({{N_{u} - N_{\bar{u}}}\over{N}}\right)_{N \rightarrow \infty} =
 \left({{N_{d} - N_{\bar{d}}}\over {N}}\right)_{N \rightarrow \infty} =
\left({{N_{s} - N_{\bar{s}}}\over {N}}\right)_{N \rightarrow \infty} =
0 \,.
}
This proves that the fundamental state is formed by pairs $u\bar{u}$
or $d\bar{d}$ or $s\bar{s}$.

These methods can be generalized easily to higher representations of
$su(3)$; the only change is in the $g$ functions (that we call source
functions): they change according to the highest weight of the
representation. The generalization to $su(n)$ is straighforward too,
but laborious; the MBA will have $(n-1)$ steps and will be
described by $(n-1)$ source functions, related by $(n-1)$ Bethe
equations.

\noindent{ \tenbf  Acknowledgements}

We are grateful to professor H. de Vega for his very useful discussions and
remarks. A
careful reading of the manuscript by professor J. Sesma is also acknowledged.
This work was partially supported by the Direcci\'{o}n General de
Investigaci\'{o}n Cient\'{\i}fica y T\'{e}cnica, Grant No PB93-0302 and
AEN94-0218

\listrefs
\end{document}